\documentclass[12pt,a4paper]{article}
\usepackage{amssymb}
\usepackage{graphicx}
\usepackage{amsmath}

\newcommand*{\cc}[1]{ \rlap{$\stackrel{*}{\phantom{#1}}$}#1 }
\newcommand{\uGamma}{\hat{\underline\Gamma}{\vphantom{\Gamma}}}
\newcommand{\mur}{{\underline{m}}}
\newcommand{\nur}{{\underline{n}}}
\newcommand{\alphar}{{\underline{a}}}
\newcommand{\betar}{{\underline{b}}}
\newcommand{\sign}{\mathop{{\rm sign}}}
\newcommand{\ctimes}{{\times \!\!\!\!\!\!\supset}}

\begin{document}

\title{Field on Poincar\'{e} group and quantum description of orientable objects}
\author{D.M. Gitman${}^a$\thanks{E-mail: gitman@dfn.if.usp.br}
    and A.L. Shelepin${}^b$\thanks{E-mail: alex@shelepin.msk.ru} \\
\\
${}^a$Instituto de F\'{\i}sica, Universidade de S\~{a}o Paulo,\\
Caixa Postal 66318-CEP, 05315-970 S\~{a}o Paulo, S.P., Brazil\\
${}^b${Moscow Institute of Radio Engineering, Electronics and Automation,}\\
Prospect Vernadskogo, 78, 117454, Moscow, Russia }
\maketitle

\begin{abstract}
We propose an approach to the quantum-mechanical description of
relativistic orientable objects. It generalizes Wigner's ideas
concerning the treatment of nonrelativistic orientable objects
(in particular, a nonrelativistic rotator) with the help of
two reference frames (space-fixed and body-fixed). A technical
realization of this generalization (for instance, in $3+1$ dimensions)
amounts to introducing wave functions that depend on elements
of the Poincar\'{e} group $G$. A complete set of transformations that
test the symmetries of an orientable object and of the embedding
space belongs to the group $\Pi =G\times G$. All such transformations
can be studied by considering a generalized regular representation of
$G$ in the space of scalar functions on the group, $f(x,z)$, that
depend on the Minkowski space points $x\in G/\mathrm{Spin}(3,1)$
as well as on the orientation variables given by the elements $z$ of
a matrix $Z\in \mathrm{Spin}(3,1)$. In particular, the field $f(x,z)$ is
a generating function of usual spin-tensor multicomponent fields.
In the theory under consideration, there are four different types
of spinors, and an orientable object is characterized by ten quantum
numbers. We study the corresponding relativistic wave equations and
their symmetry properties.
\end{abstract}

\section*{Introduction}

The problem of a quantum-mechanical description of orientable objects is not
one of the issues that are frequently encountered in the literature,
although the general approach to such a description cannot
be considered as being completely formulated. It is well-known that for a
quantum-mechanical description of a point-like spinless particle in an
$n$-dimensional (pseudo)Euclidean space it is sufficient to use one wave
function that depends on space-time coordinates $x^{\mu }$ alone. A complete
description of orientable objects requires some additional coordinates. For
example, in order to determine the exact localization of a rigid body in a
$3$-dimensional space one needs to assign three coordinates that determine the
position of its mass center, as well as three angles that determine the
orientation. It is natural to consider a quantum-mechanical description of
such orientable objects as being achieved by an introduction of wave
functions depending not only on the $n$ coordinates of the mass center, but
also on some auxiliary variables that describe the orientation. In the known
examples (a spinning particle and a non-relativistic rigid rotator), the
orientation is usually taken into account by an introduction of
multi-component wave functions depending on the space-time coordinates
$x^{\mu }$. As to the first example, the construction and classification of
such functions is largely due to the theory of representations of the Poincar\'{e}
and Lorentz groups, rather than due to the well-formulated theory that
describes spinning particles as particles that possess orientation.

Until recently, the only example of a well-developed physical theory in
which wave functions depend on orientation (and only on orientation) has
been the theory of the above mentioned rigid rotator, constructed by Wigner,
Casimir and Eckart back in the 1930's (see \cite{BieLo81} for references and
historical remarks). One reference frame (laboratory, space-fixed,
\emph{s.r.f. in what follows}) is assigned with the surrounding objects, while
another one (molecular, body-fixed, \emph{b.r.f. in what follows}) is
assigned with the rotating body. Correspondingly, there are two sets of
operators of angular momentum: those of s.r.f. (left generators of the
rotation group $\hat{J}_{k}^{L}$), and those of the b.r.f. (right generators
of the rotation group $\hat{J}_{k}^{R}$). The interpretation of $\hat{J}_{k}^{R}$
as projections of momentum in the b.r.f. belongs to Wigner and
Casimir (1931), and basically lays the foundation of the theory of molecular
spectra. Mathematically, the construction of a theory of a non-relativistic
rigid rotator is a construction of a filed on the group $G=SO(3)\thicksim SU(2)$,
see below.

It should be noted that the description of relativistic spinning particles
can be reformulated in terms of one scalar field depending on the space-time
coordinates $x^{\mu }$ as well as on some auxiliary continuous
variables that describe spin. Such a reformulation has a long history. At
the end of 1940's and the beginning of\ 1950's, independently by various
authors \cite{GinTa47,BarWi48,Yukaw50,Shiro51}, mainly in connection with a
construction of relativistic wave equations (RWE), were introduced some fields
depending not only on $x^{\mu }$ but also on a certain set of spinning variables.
A systematic treatment of these fields as fields on homogenous spaces of the
Poincar\'{e} group was made by Finkelstein \cite{Finke55} in 1955. He also
presented a classification and explicit constructions of homogenous spaces
of the Poincar\'{e} group which contain the Minkowski space that is a homogenous
space of the latter group. In 1964, Lur\c{c}at \cite{Lurca64} suggested to construct
a quantum theory on the whole Poincar\'{e} group, instead of the Minkowski space.

In 70-90, the ideas of constructing fields in various homogenous spaces of
the Poincar\'{e} group, gained a certain development, in particular, in the
papers \cite{BacKi69,Kihlb70,BoyFl74,Arodz76,Tolle78,Tolle96,Drech97,Hanni97}.
Properties of various spaces were considered, as well as some
possibilities for introducing interactions in the spin phase-space and
possibilities for constructing Lagrangian formulations.
Restrictions were studied to be imposed on scalar fields by using a choice
of a homogenous space. Thus, the authors of \cite{BacKi69} arrived at the
conclusion that the minimal dimension of a homogenous space suitable for a
description of integer and semi-integer spins equals to eight. In the papers
\cite{GitSh01,BucGiS02}, we have developed a general approach to
constructing fields on groups of motions in Euclidean and pseudo-Euclidean
spaces, elaborating in detail the cases of $2,3$- and $4$-dimensions. In that
approach, a scalar field on the Poincar\'{e} group is a generating function
of usual multicomponent fields. In particular, it has been demonstrated
that, as distinct from the case of scalar fields on homogenous spaces, a
field on a group as a whole is closed with respect to discrete
transformations. The task of constructing RWE in that approach looks
especially natural, since it is intimately related with a classification of
scalar functions on a group.

However, until now there has been no clear understanding of a connection
between the approach that describes nonrelativistic orientable objects
(\`{a} la Wigner) and the one that describes spinning particles in terms of
multicomponent functions. In the present article, we look at both problems
from a viewpoint that suggests a universal approach to a quantum-mechanical
description of orientable objects on a basis of the theory of
representations of spatial groups, using, at the same time, some physical
concepts that bear their origin in the intuitively obvious nonrelativistic
theory of orientable objects. Such a treatment of relativistic spinning
particles allows a natural appearance and interpretation of constructions
that have been frequently introduced by hand, in a heuristic manner.

The first observation that plays a key role in the suggested approach is the
following (we further make a reference to the easily imaginable
$3$-dimensional Euclidean case): a description of an orientable object is
completely determined by the positions of the corresponding b.r.f. with
respect to some s.r.f., which means that there exists a technical
possibility to describe these positions by elements of the corresponding
transformation group (namely, the (pseudo)Euclidean group of motions $G$);
see examples below. Thus, we believe that quantum mechanical description of
orientable objects can be performed by scalar wave functions $f(g)$ on the
corresponding group of motions, $g\in G$.

The following step is a realization of the fact that testing the symmetry of
a space that contains an orientable object is related to the behavior of
wave functions with respect to the transformations of the s.r.f., which we
further call left transformations, $g_{l}\in G$, and which belong to the
same group $G$, whereas testing the symmetry properties of the orientable
object itself is related to the behavior of wave functions with respect to
the transformations of the b.r.f., which we further call right
transformations, $g_{r}\in G$, and, which, once again, belong to $G$. We
thus arrive at the necessity of studying the behavior of the wave function
describing an orientable object under the action of transformations being
the direct product of left and right transformations, i.e., under the action
of the group $\Pi =G\times G$.

Finally, an issue that concludes the construction is the fact that a
complete classification of the wave functions $f(g)$ of an orientable object
under the action of transformations that belong to $\Pi $ can be extracted
from the study of the so-called generalized regular representation (GRR)
\[
\mathbb{T}( g_{l};g_{r}) f(g) =f(g_{l}^{-1}g~g_{r}) ,
\]
while the right and left generators of the group $G$ in this representation
are interpreted as ones related to the above-defined right and left
transformations.

The present article begins with a consideration of some simple cases that
describe orientable objects in $2$ and $3$ dimensions. In these cases, the
use of visually obvious concepts allows one to present a convincing
demonstration of the possibility to realize the above-declared program. We
next present our approach in its general setting, and finally proceed to a
detailed treatment of the physically most interesting description of
orientable objects in the Minkowski space, in which a description of the
position of an orientable object requires to indicate, besides the four
coordinates $x_{\mu }$, also its orientation with respect to the laboratory
reference frame. The latter is given by a pseudoorthogonal matrix
$V\in SO(3,1)$. Thus, the orientable object in the Minkowski space is described by
a pair $(x,V)$. An element of the group of motions
$M_{0}(3,1)=T(4)\ctimes SO(3,1)$ is given by the same pair
$(a,\Lambda )$, where $a$ is a translation, and $\Lambda $ is a rotation.
It is easy to verify that under changes of the laboratory and body-fixed
reference frames, respectively, we have
\[
(x',V')=(a,\Lambda)^{-1}(x,V), \qquad (x',V')=(x,V)(a,\Lambda).
\]

In what follows, we consider a generalized regular representation
being a representation in the space of functions on the Poincar\'{e} group.
An important (not only technically) question is a parametrization of the
matrices $V$. Using the homomorphism $SL(2,C)={\rm Spin}(3,1)\sim SO(3,1)$, we
consider functions $f(x,z)$ of space-time coordinates $x$ and complex-valued
spinor variables $z$ being the elements of a matrix $Z\in SL(2,C)$.

The field $f(x,z)$ is a generating function of the usual spin-tensor
multicomponent fields (they appear as coefficients of the power expansion
with respect to $z$) and admits a number of symmetry operations.

The maximal set of commuting operators (their number being equal to that of
the group parameters) in the space of functions consists of the Casimir
operators and of the (equal in number) left and right generators.
Functions $f(x,z)$ on the 3+1 Poincar\'{e} group depend on ten parameters.

In the conventional description of relativistic particles (in $3+1$ dim.) in
terms of spin-tensor fields, based, in particular, on a classification of
Poincar\'{e} and Lorentz group representations, there appear 8 particle
characteristics (quantum numbers): 2 numbers $j_{1}$ and $j_{2}$ that label
Lorentz group representations and 6 numbers related to the Poincar\'{e} group,
those are the mass $m$, spin $s$ (Casimirs of the group), and 4 numbers,
which are eigenvalues of some combinations of left generators of the group.
In particular, the latter 4 numbers can be some components of the momentum
and a spin projection.
The proposed description of relativistic orientable objects is based on a
classification of group representations of transformation of both s.r.f. and
b.r.f., and the orientable object is characterized already by 10 quantum
numbers, which are related to the maximal set of commuting operators.

We introduce the concept of a symmetry of a field on a group and classify
the symmetries of a field on the Poincar\'{e} group. In particular, it turns
out that the discrete transformations ($C,P,T$) correspond to involutive
automorphisms of the group and are reduced to a complex conjugation and to a
change of arguments of scalar functions $f(x,z)$.

The orientation variables $z$, as well as the corresponding decomposition
coefficients, have two types of indices, related to the above-mentioned left
and right transformations. The ``left'' indices related to changes of the
s.r.f. (Lorentz transformations) are the usual (vector or spinor) indices,
whereas ``right'' indices are related to changes of the b.r.f.
Consequently, in contrast with nonrelativistic theory (in which there is
only one type of spinors) and usual relativistic theory (in which there are
two types of spinors), in a relativistic theory of oriented objects there
are four types of spinors with different rules of transformations.

Using left-invariant (i.e., Lorentz invariant) differential operators of
first order with respect to the variables $x,z$, in the corresponding
eigenvalue problems, we arrive to RWE in the theory under consideration.
Reducing the general RWE to the space of $f(x,z)$ being polynomials of a fixed order
$2s$ in $z$, we may obtain, in particular, all the known types of RWE for spin $s$
in spin-tensor representation.
We study RWE and their symmetries with respect to the symmetries of functions
on the Poincar\'{e} group (left and right transformations, outer
automorphisms). In such a way, there appear different types of RWE
(reducible from the point of view of left transformations) which play
important role in physics and which often have to be constructed ``by hands''.
For example, here we naturally obtain eight-component relativistic wave
equation for particle of spin 1/2, which was derived in course of a
canonical quantization from Berezin-Marinov action (see \cite{Git}), and which allows
one to avoid difficulties with infinite number of negative energy levels to
construct a consistent version of relativistic quantum mechanics of
noninteracting spinning particles.

\section{Two dimensional Euclidean case}

Let us consider a two-dimensional Euclidean space with a Cartesian reference
frame, given by an orthonormalized basis $\boldsymbol{e}_{i},\ i=1,2$
(space-fixed reference frame, or s.r.f.).
Suppose we would like to describe some orientable
object in the s.r.f. (as such an object, one can imagine a two-dimensional
solid body). To this end, we attach to the orientable object an additional
Cartesian frame (body-fixed reference frame, or b.r.f.), given by an
orthonormalized basis $\boldsymbol{\xi}_{i},\ i=1,2$. Then, the orientable object
is described by a position $\boldsymbol{x}=(x^{i},\;i=1,2)$ of the
b.r.f. origin with respect to the s.r.f., and by an angle $\theta $ between
the corresponding axis of the s.r.f. and b.r.f.
Therefore, as coordinates of the orientable object, we choose the pair
$\boldsymbol{x},\theta $ .

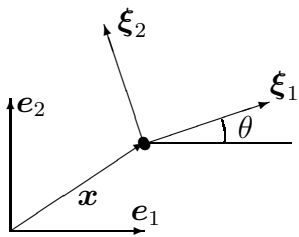
\begin{picture}(360,110)
\put(10,10){\vector (1,0){50}}
\put(10,10){\vector (0,1){50}}
\put(12,55){$\boldsymbol{e}_2$}
\put(55,14){$\boldsymbol{e}_1$}

\put(60,43){\vector (3,1){47}}
\put(60,43){\vector (-1,3){15}}
\put(50,86){$\boldsymbol{\xi}_2$}
\put(107,62){$\boldsymbol{\xi}_1$}
\put(60,43){\line (1,0){55}}

\put(10,10){\vector (3,2){50}}
\put(60,43){\circle*{4}}
\put(59.7,43.6){\circle*{4}}
\put(60.3,42.4){\circle*{4}}
\put(35,20){$\boldsymbol{x}$}
\qbezier(90,43)(91,47)(89,52)
\put(95,45){$\theta$}
\end{picture}

We will now look to two types of transformations.
Translations of the s.r.f. origin by a vector $\boldsymbol{a}$ with its
rotations by an angle $\phi $ form the first type; in what follows, we
call them left transformations. Translations of
the b.r.f. origin by a vector $\boldsymbol{b}$ with its rotations by an angle
$\psi$ form another type; in what follows, we call them right transformations.

The coordinates $\boldsymbol{x}$,$\theta $ of the orientable object are
changed to $\boldsymbol{x}'$,$\theta '$ under the left
transformations:
\begin{align}
& {x^{1}}^{\prime}=(x^{1}-a^{1})\cos\phi+(x^{2}-a^{2})\sin\phi\,,  \notag \\
& {x^{2}}^{\prime}=(x^{2}-a^{2})\cos\phi-(x^{1}-a^{1})\sin\phi\,,  \notag \\
& \theta^{\prime}=\theta-\phi\,,  \label{e2d.3}
\end{align}
where $a^{1}$ and $a^{2}$ are components of the translation $\boldsymbol{a}$ in
the s.r.f.. The left transformations form the group $M(2)=T(2)\ctimes SO(2)$
of motions of the two-dimensional Euclidean space.

Under the right transformations, the coordinates $\boldsymbol{x}$,$\theta $
are changed as follows:
\begin{align}
& {x^{1}}^{\prime}=x^{1}+b^{1}\cos\theta-b^{2}\sin\theta\,,  \notag \\
& {x^{2}}^{\prime}=x^{2}+b^{2}\cos\theta+b^{1}\sin\theta\,,  \notag \\
& \theta^{\prime}=\theta+\psi\,,  \label{e2d.4}
\end{align}
where $b^{1}$ and $b^{2}$ are components of the translation $\boldsymbol{b}$ in the b.r.f..
The right transformations also form the group $M(2)=T(2)\ctimes SO(2)$.

Consider now transformations realized as combinations of all possible right
and left transformations. The general transformations form a group, which is
the direct product $\Pi =M(2)\times M(2)$. The general transformations act
on the coordinates $\boldsymbol{x}$,$\theta $ as follows:
\begin{align}
& {x^{1}}^{\prime}=(x^{1}-a^{1}+b^{1}\cos\theta-b^{2}\sin\theta)\cos\phi
+(x^{2}-a^{2}+b^{2}\cos\theta+b^{1}\sin\theta)\sin\phi\,,  \notag \\
& {x^{2}}^{\prime}=(x^{2}-a^{2}+b^{2}\cos\theta+b^{1}\sin\theta)\cos\phi
-(x^{1}-a^{1}+b^{1}\cos\theta-b^{2}\sin\theta)\sin\phi\,,  \notag \\
& \theta^{\prime}=\theta-\phi+\psi\,.  \label{e2d.10}
\end{align}

The left, right, and general transformations can be described, in a
convenient manner, with the help of a $3\times 3$-matrix function
$g\left(\boldsymbol{r},\varphi \right) $ of the variables $\boldsymbol{r}$
and $\varphi$. This matrix function has the form
\begin{align}
& g\left( \boldsymbol{r},\varphi\right) =\left(
\begin{array}{cc}
V\left( \varphi\right) & \boldsymbol{r} \\ \mathbf{0} & 1
\end{array}
\right) ,
\quad
\boldsymbol{r}=\left(
\begin{array}{c}
r^{1} \\ r^{2}
\end{array}
\right) ,\notag \\
& V\left( \varphi\right) =\Vert v_{\;k}^{i}\left( \varphi\right)
\Vert=\left(
\begin{array}{cc}
\cos\varphi & -\sin\varphi \\
\sin\varphi & \cos\varphi
\end{array}
\right) ,
\quad
\mathbf{0}=\left(
\begin{array}{cc}
0 & 0
\end{array}
\right) .  \label{e2d.11}
\end{align}

Let the initial position of the orientable object be $\boldsymbol{x}$,$\theta .$
Then, the mutual orientation of the bases $\boldsymbol{e}_{i}$ and
$\boldsymbol{\xi}_{i}$ is given by a $2\times 2$ matrix $V\left( \theta \right) $,
\begin{align}
\boldsymbol{\xi}_{k}=\boldsymbol{e}_{i}v_{\;k}^{i}\left( \theta\right) \,,\;
V\left(\theta\right) =\Vert v_{\;k}^{i}\left( \theta\right) \Vert=\left(
\begin{array}{cc}
\cos\theta & -\sin\theta \\
\sin\theta & \cos\theta
\end{array}
\right) .  \label{e2d.1}
\end{align}
One can verify that the left transformations (\ref{e2d.3}),
the right transformations (\ref{e2d.4}),
and the general transformations (\ref{e2d.10}) respectively, can be written as
\begin{align}
&g\left( \boldsymbol{x}^{\prime},\theta^{\prime}\right)
=g^{-1}\left( \boldsymbol{a},\phi\right) g\left( \boldsymbol{x},\theta\right) \,,  
\nonumber \\
&g\left( \boldsymbol{x}^{\prime},\theta^{\prime}\right)
=g\left( \boldsymbol{x},\theta\right) g\left( \boldsymbol{b},\psi\right) \,,  
\nonumber \\
&g\left( \boldsymbol{x}^{\prime},\theta^{\prime}\right)
=g^{-1}\left( \boldsymbol{a},\phi\right) g\left( \boldsymbol{x},\theta\right) g\left( \boldsymbol{b},\psi\right)
\,.  \label{e2d.5}
\end{align}

Relation (\ref{e2d.5}) implies
\begin{align}
g_{\alpha m}\left( \boldsymbol{x}^{\prime},\theta^{\prime}\right)
=\sum_{\beta,c}g_{\alpha\beta}^{-1}\left( \boldsymbol{a},\phi\right)
g_{\beta c}\left( \boldsymbol{x},\theta\right) g_{cm}\left( \boldsymbol{b},\psi\right)\,
=\sum_{\beta,c}\mathbf{t}_{\alpha m|\beta c}\left( \boldsymbol{a},\phi;\boldsymbol{b},\psi\right)
g_{\beta c}\left( \boldsymbol{x},\theta\right) \,,
\label{e2d.14}
\end{align}
where
\begin{equation*}
\mathbf{t}_{\alpha m|\beta c}( \boldsymbol{a},\phi;\boldsymbol{b},\psi)
=g_{\alpha\beta}^{-1}( \boldsymbol{a},\phi) g_{cm}( \boldsymbol{b},\psi) \,,
\end{equation*}
or, introducing the composed and ordered indices
$A=\left( \alpha m\right),\;B=\left( \beta c\right)$, we have
\begin{align}
g_{A}\left( \boldsymbol{x}^{\prime},\theta^{\prime}\right) =\sum_{B}\mathbf{t}%
_{A|B}\left( \mathbf{a},\phi;\boldsymbol{b},\psi\right) g_{B}\left( \boldsymbol{x}%
,\theta\right) \,.  \label{e2d.8}
\end{align}

One can consider (\ref{e2d.14}) as a vector representation of the group $\Pi$ in
the space of ``vector functions'' $g_{B}\left( \boldsymbol{x},\theta \right)$.
For $\boldsymbol{b}=\psi =0$, this is a left vector representation of the group
$M(2)$, and for $\boldsymbol{a}=\phi =0$ this is a right vector representation
of the same group.

Let us define another representation of the group $\Pi $ that acts on
scalar functions $f\left( g\left( \boldsymbol{x},\theta \right) \right) $ of
the vectors $g_{B}\left( \boldsymbol{x},\theta \right) ,$
\begin{align}
& \mathbb{T}\left( \boldsymbol{a},\phi;\boldsymbol{b},\psi\right) f\left( g\left(
\boldsymbol{x},\theta\right) \right) =f^{\prime}\left( g\left( \boldsymbol{x},
\theta\right) \right) =f\left( g\left( \boldsymbol{x}^{\prime},\theta^{\prime}\right) \right)
\notag \\
& \,=f\left( \mathbf{t}\left( \boldsymbol{a},\phi;\boldsymbol{b},\psi\right) g\left(
\boldsymbol{x},\theta\right) \right) =f\left( g^{-1}\left( \boldsymbol{a},
\phi\right) g\left( \boldsymbol{x},\theta\right) g\left( \boldsymbol{b},\psi\right)
\right) .  \label{e2d.9}
\end{align}
One can verify that (\ref{e2d.9}) is actually a representation. In what
follows, it is called the regular representation. For this representation,
the left (for $\boldsymbol{b}=0,\psi =0$) and right (for
$\boldsymbol{a}=0,\phi =0)$ regular representations of the group $M(2)$ are
subrepresentations. Thus, the generators of the group $\Pi $ in the regular
representation consist of all the generators of the right and left regular
representations of the group $M(2)$.

An expansion of the left or right regular representation contains (with
accuracy up to an equivalence) all irreps of the group $M(2)$.

For the left and right generators (corresponding to the parameters $a_{k},
\phi$ and $b_{k}, \psi$), we have
\begin{align}  \label{e2d.15}
& \hat p_{k}=-i\partial_{k},\quad\hat J=\hat L+\hat S, \\
& \hat p_{k}^{R}=iV_{\;\;k}^{i} \partial_{i},\quad\hat J^{R}=-\hat S,
\label{e2d.15a}
\end{align}
where
\begin{equation*}
\hat L=i(x^{1}\partial_{2} -x^{2}\partial_{1} ), \quad
\hat S={-i} \frac{\partial}{\partial\theta}.
\end{equation*}
An invariant measure on the group has the form
$d\mu(x,\theta)=(4\pi)^{-1}dx_{1}dx_{2}d\theta$,
where integration is taken over the manifold ${R}^{2}\times S^{1}$.

The group $M(2)$ is a three-parameter one; therefore, a maximal set of
commuting operators includes three operators. In accordance with the general
theory, this set is formed by the Casimir operators plus the equal number of
left and right generators. We can select this set as the Casimir operator
$\hat{\boldsymbol{p}}^{2}=\hat p_{1}^{2}+\hat p_{2}^{2}=(\hat
p_{1}^{R})^{2}+(\hat p_{2}^{R})^{2}$, plus the left generator $\hat J$ and
the right generator $\hat S$, i.e., the operators of squared momentum, total
momentum and intrinsic momentum.

Let us return to relation (\ref{e2d.9}) that determines the action of the
group $M(2)\times M(2)$. In the general case, irreps of this group are
characterized by two (generally) different numbers. In the case under consideration,
the representation $T_{\Pi }(g,h)$ acts on functions of three variables
only, and in this space one can only construct a part of irreps of
$M(2)\times M(2)$, the Casimir operators of the
subgroups $M(2)$ being identical, ${\hat{p}_{L}}^{2}={\hat{p}_{R}}^{2}$.
At the same time, the projections $J_{L}$ and $J_{R}$ may take various values.

Note that there is a certain inequivalence of left and right transformations
in a given interpretation. If we consider a point like object without an orientation,
we cannot relate it to any definite b.r.f., and, consequently, we cannot determine
right transformations. In this case, the intrinsic momentum equals to zero, so it
is sufficient to examine only the left transformations,
the related two quantum numbers ($p_{1},p_{2}$ or $\boldsymbol{p}^{2},\,J=L$)
completely characterize a non-orientable object.

Two dimensions, however, do not allow one to perceive, in full measure, the
peculiarity inherent in the description of orientable objects, because the
subgroup of rotations $SO(2)$ that describes the intrinsic momentum is
commutative. For commutative groups, left and right generators obviously
coincide. In the given case, due to the commutativity, the operator of intrinsic
momentum in the b.r.f. (i.e., the left generator of the
rotation subgroup) coincides, up to sign, with the operator of intrinsic momentum in
the s.r.f. (i.e., identical with one of the summands of
the left generator): $\hat{J}_{L}=\hat{L}+\hat{S}$, $\hat{S}=-\hat{J}_{R}$.
In higher-dimensional spaces, this is not the case.

Therefore, prior to considering the case of four dimensions and the Poincar\'{e}
group, we are going to construct a theory of a three-dimensional rigid
rotator (which provides a description of the \textit{intrinsic} momentum of
a rigid body) in terms of the right and left transformations of the
non-commutative group $SU(2)$.

\section{ Non-relativistic three-dimensional rotator}

\subsection{General}

To describe a rigid rotator (gyroscope), we will use two
orthonormalized reference frames: s.r.f.  $\{\boldsymbol{e}_{i},\ i=1,2,3\}$
and b.r.f. $\{\boldsymbol{\xi}_{k},\ k=1,2,3\}$,
\begin{align}  \label{rot.1}
\boldsymbol{\xi}_{k}= \boldsymbol{e}_{i} v^{i}_{\;k} .
\end{align}
The scalar product $(\boldsymbol{e}_{i},\boldsymbol{e}_{j})=\delta _{ij}$ ,
and, therefore, the elements of the matrix $V=\Vert v_{\;k}^{i}\Vert $
satisfy the condition $v_{\;k}^{i}v_{\;l}^{i}=\delta _{kl}$, i.e., the
matrix $V$ is orthogonal, $V^{T}=V^{-1}$.

Thus, the orientation of a three-dimensional rotator is determined by a
$3\times 3$ orthogonal matrix, $V\in O(3)$, composed of the coefficients of a
re-decomposition of the bases (s.r.f. and b.r.f.).
If both systems $\{\boldsymbol{e}_{i}\}$ and $\{\boldsymbol{\xi}_{k}\}$
are right or left, the matrix $V\in SO(3)$ depends on three
real-valued parameters, which can be chosen as the Euler angles:
\begin{align}
&V=
\left(\begin{array}{ccc}
\cos \phi & -\sin \phi & 0\\
\sin \phi & \cos \phi & 0 \\
0 & 0 & 1 \\
\end{array}\right)
\left(\begin{array}{ccc}
\cos \theta & 0 & \sin \theta \\
0 & 1  & 0 \\
-\sin \theta & 0 & \cos \theta
\end{array}\right)
\left(\begin{array}{ccc}
\cos \psi & -\sin \psi & 0\\
\sin \psi & \cos \psi & 0 \\
0 & 0 & 1 \\
\end{array}\right) =
\nonumber
\\
&\left(
\begin{array}{ccc}
\cos\phi\cos\psi\cos\theta -\sin\phi\sin\psi   & -\sin\phi\cos\psi-\cos\phi\sin\psi\cos\theta\; & \cos\phi\sin\theta \\
\sin\phi\cos\psi\cos\theta +\cos\phi\sin\psi\; & \cos\phi\cos\psi-\sin\phi\sin\psi\cos\theta   & \sin\phi\sin\theta \\
-\cos\psi\sin\theta                              & \sin\psi\sin\theta                             & \cos\theta  \\
\end{array}
\right).
\label{rot.2}
\end{align}

We now examine \textit{two kinds of transformations}: rotations of the
space-fixed and body-fixed reference frames.
It is obvious that in the first case the quantities $\{\boldsymbol{e}_{i}\}$
transform as vector components, whereas in the second case they remain
intact, i.e., they are scalars with respect to internal transformations
(rotations of the body). On the contrary, the quantities $\{\boldsymbol{\xi}_{k}\}$
are scalars with respect to external transformations (rotations of
s.r.f.), whereas they are vector components with respect to internal transformations.

The elements $v_{\;k}^{i}$ specify the body orientation with respect to s.r.f.,
and we can consider $v_{\;k}^{i}$ as coordinate set of rotator.
Let us describe the transformations of the set
$v_{\;k}^{i}$ (the first index being ``external'', the second being ``internal'')
under the rotations of the s.r.f. and b.r.f..

In the matrix notations, we can present (\ref{rot.1}) as $\xi=eV$. A rotation
of the s.r.f. $e'=e\Lambda$ yields $\xi=eV=e'\Lambda^{-1}V=e'V'$, whence
\begin{equation}
V'=\Lambda^{-1}V.\label{rot.left}
\end{equation}
A rotation of the b.r.f. $\xi'=\xi\underline{\Lambda}$ yields
$\xi =\xi'\underline{\Lambda}^{-1}=eV$, whence
\begin{equation}
V^{\prime}=V\underline{\Lambda},\label{rot.right}
\end{equation}
and, therefore, external transformations correspond to left-multiplication,
whereas internal ones correspond to right-multiplication.

The general transformations have the form
\begin{align}  \label{rot.gen}
V^{\prime}=\Lambda^{-1} V \underline{\Lambda} .
\end{align}
By analogy with the matrix $V$ that determines the orientation, the rotation
matrices $\Lambda $ and $\underline{\Lambda }$ are parameterized by three
Euler angles. In the representation (\ref{rot.gen}), generators are given by
the standard $3\times 3$ matrices. In addition, the matrices of generators
of transformations (\ref{rot.left}), (\ref{rot.right}) have the same form
(however, their action is different, being related to left- and right-multiplication).

To find generators of an arbitrary irrep of $SO(3)$, one has
to examine representations in the space of functions on the group, i.e.,
functions $f(\phi,\psi,\theta)$ of the rotator orientation.

The left regular representation $T_{L}(g)$ acts in the space of functions
$f(q)$, $q=q(\phi ,\psi ,\theta )\in SO(3)$, on the group as follows:
\begin{align}  \label{rot.5}
T_{L}(g)f(q)=f^{\prime}(q)=f(g^{-1}q),\;g\in G,
\end{align}
which corresponds to a change of the s.r.f.; see
(\ref{rot.left}); whereas the right regular representation $T_{R}(h)$ acts
in the same space as follows:
\begin{align}
T_{R}(h)f(q)=f^{\prime}(q)=f(qh),\quad h\in G,  \label{rot.6}
\end{align}
which corresponds to a change of the b.r.f.; see
(\ref{rot.right}). The decomposition of the left (and right) regular
representation contains any irrep of the group.

Each set of the left and right transformations forms the group $SO(3)$.
Since these two transformation sets commute with each other, we can consider
them as the direct product $\Pi =SO(3)\times SO(3)$. The transformations from
$\Pi $ act in the space of functions depending on three parameters
(on the rotator orientation) as follows:
\begin{align}
T_{\Pi}(g,h)f(q)=f(g^{-1}qh)=f^{\prime}(q) .
\end{align}
It is obvious that the generators of $\Pi $ in this representation
consist of the generators of the subgroups $SO(3)$ (\ref{rot.5}) and (\ref{rot.6}).

For generators that correspond to the one-parameter subgroup $\omega (t)$
in the left and right regular representations, we have
\begin{align}  \label{rot.7}
\hat J_{\omega}f(q)= -i\lim_{t\to0} \frac{f(\omega^{-1}(t)q) - f(q)}{t},
\qquad\hat I_{\omega}f(q)= -i\lim_{t\to0} \frac{f(q\omega(t)) - f(q)}{t},
\end{align}
the multiplier $i$ provides the hermiticity of the generators.
Accordingly, the operators of finite transformations corresponding to these
one-parameter subgroups are given by
\begin{equation*}
T_{L}(\omega(t))=\exp(i\hat J_{\omega}t), \qquad T_{R}(\omega(t))=\exp(i\hat
J_{\omega}t).
\end{equation*}

Let us denote by $\phi (t),\psi (t)$, and $\theta (t)$ the Euler angles of an
element $\omega ^{-1}(t)q$. Then,
\begin{equation*}
\hat J_{\omega}f(q)= -i\left. \frac{df(\omega^{-1}(t)q)}{dt}\right| _{t\to
0}= -i \left( \frac{df}{d\phi}\phi^{\prime}(0)+\frac{df}{d\theta}
\theta^{\prime}(0)+\frac{df}{d\psi}\psi^{\prime}(0) \right) .
\end{equation*}
Having denoted by $\phi (t),\psi (t)$, and $\theta (t)$ the Euler angles of the
element $q\omega (t)$, we obtain an analogous formula for the right
generators.

Let us choose the one-parameter subgroups as follows:
\begin{equation*}
\omega_{1}=\left(
\begin{array}{ccc}
1 & 0 & 0 \\
0 & \cos t & \sin t \\
0 & -\sin t & \cos t
\end{array}
\right) , \quad\omega_{2}=\left(
\begin{array}{ccc}
\cos t & 0 & -\sin t \\
0 & 1 & 0 \\
\sin t & 0 & \cos t
\end{array}
\right) , \quad\omega_{3}=\left(
\begin{array}{ccc}
\cos t & \sin t & 0 \\
-\sin t & \cos t & 0 \\
0 & 0 & 1
\end{array}
\right) .
\end{equation*}
The transformations $\omega _{k}^{-1}(t)q$ correspond to rotations about
the axes $\boldsymbol{e}_{k}$, whereas $q\omega _{k}(t)$ correspond to rotations about
$\boldsymbol{\xi}_{k}$. Direct calculations yield the following expressions
for the generators of the s.r.f. rotations
\begin{align}
&\hat J_1=-i\left(  \frac{\cos\phi}{\sin\theta}\frac{\partial}{\partial\psi} - \sin\phi\frac{\partial}{\partial\theta} -
\cos\phi \cot\theta \frac{\partial}{\partial\phi}\right),
\nonumber 
\\
&\hat J_2=-i\left( \frac{\sin\phi}{\sin\theta}\frac{\partial}{\partial\psi} + \cos\phi\frac{\partial}{\partial\theta} -
\sin\phi \cot\theta \frac{\partial}{\partial\phi}\right),
\nonumber 
\\
&\hat J_3=-i\frac{\partial}{\partial\phi},
\label{rot.J3}
\end{align}
and b.r.f. rotations
\begin{align}
&\hat I_1=-i\left(  \frac{\cos\psi}{\sin\theta}\frac{\partial}{\partial\phi} - \sin\psi\frac{\partial}{\partial\theta} -
\cos\psi \cot\theta \frac{\partial}{\partial\psi}\right),
\nonumber 
\\
&\hat I_2= i\left( \frac{\sin\psi}{\sin\theta}\frac{\partial}{\partial\phi} + \cos\psi\frac{\partial}{\partial\theta} -
\sin\psi \cot\theta \frac{\partial}{\partial\psi}\right),
\nonumber 
\\
&\hat I_3= i\frac{\partial}{\partial\psi}.
\label{rot.I3}
\end{align}

It is easy to see that all the right generators commute with all the left
generators,
\[
[\hat J_{i},\hat I_{k}]=0, \quad
[\hat J_{i},\hat J_{k}]=i\epsilon^{ikl}\hat J_{l}, \quad
[\hat I_{i},\hat I_{k}]=i\epsilon^{ikl}\hat I_{l}.
\]
This follows from the associativity of the group multiplication: in the
product $g^{-1}qh$ the result does not depend on whether one multiplies
first from the right or from the left.

The quantities $\hat{I}_{k}$ remain the same with a change of the
s.r.f., and, therefore, they are three ``external''
(coordinate) scalars; however, with a change of b.r.f. they
transform as vector components. That is, $\hat{J}_{k}$ and $\boldsymbol{e}_{k}$
are ``external'' vectors and ``internal'' scalars; $\hat{I}_{k}$ and
$\boldsymbol{\xi}_{k}$ are ``external'' scalars and ``internal'' vectors. The
quantities $v_{\;k}^{i}$ possess one ``external'' and one ``internal'' index.

Let us construct a minimal set of commuting operators in the space of functions $f(\phi ,\psi,\theta )$.
The algebra of operators $\hat{I}_{k}$ has the same commutation
relations as the algebra of operators $\hat{J}_{k}$, and therefore the
standard results of the angular momentum theory are immediately valid for
them. We obtain rotation multiplets of dimension $2I+1$, where $I$ is the
integer or half-integer maximal value of projection $K=I_{3}$ to the fixed
axis $\boldsymbol{\xi _{3}}$, the squared value of the momentum being $\hat{I}^{2}=I(I+1)$.
The value of total momentum does not depend on the choice of axes
(which can be verified by using the explicit form of generators),
\[
J(J+1)=\hat J^{2}=\hat J_{1}^{2}+\hat J_{2}^{2}+\hat J_{3}^{2}
=\hat I_{1}^{2}+\hat I_{2}^{2}+\hat I_{3}^{2}=\hat I^{2}=I(I+1),
\]
and, therefore, the quantum numbers $I$ and $J$ must be identical. The set
of quantities that can be measured simultaneously (commuting operators) and
characterize the states of a rotator is
\begin{align}
\hat J^{2}=\hat I^{2}, \quad \hat J_{3},\quad\hat I_{3}.
\label{rot.10}
\end{align}
Eigenvalues of these operators are $J(J+1)$, $M=-J,-J+1,\ldots ,J$, and
$K=-J,-J+1,\ldots ,J$. Thus, quantum states of a rotator $|J\,MK\rangle$ are
uniquely determined by the momentum $J$ and its two projections: $M$ to a
space-fixed axis, and $K$ to a body-fixed axis.
The dimension of the multiplet for a given value of $J$ obviously equals to
$(2J+1)^{2}$. An explicit form of the states $|J\,M\,K\rangle $ is given by
the so-called Wigner D-functions, being the matrix elements of the
irreps $T_{J}(g)$ of the group $SO(3)$.

Wave functions that do not depend of $\psi $ are eigenfunctions of
$\hat{I}_{3}$ with the eigenvalue $K=0$. In addition, the operators
$\hat{J}^{k}$ (\ref{rot.J3}) acquire the form of the ``usual''
operators of intrinsic momentum for a non-orientable point particle, which
depend only on the two angles $\theta $ and $\phi $.
Such states are $|J\,M\rangle =|J\,M\,0\rangle $.

For a given $J$, the action of operators $\hat{J}_{k}$ on $(2J+1)$ states
$|J\,M\,K\rangle$ for a fixed $K$ yields a linear combination of states with
the same $K$; a similar fact is valid with respect to the action of $\hat {I}_{k}$
with a fixed projection $M$. The action of $\hat{J}_{k}$ and $\hat {I}_{k}$
on a matrix composed of $|J\,M\,K\rangle$ can be, therefore,
represented as the (respectively, left- and right-) multiplication by
$(2J+1)\times(2J+1)$ matrices of generators in this representation.

\subsection{Symmetries}

The Hamiltonian of a rotator, $\hat{H}$,
due to spatial isotropy, cannot depend explicitly on the orientation of the
rotator, and is, therefore, an ``external'' scalar. That is, the only
combination of left generators $\hat{J}_{k}$, on which it can depend, is the
Casimir operator $\hat{J}^{2}$. However, it can also be a function of the
operators $\hat{I}_{k}$, which are ``external'' scalars as well.

Consider stationary states of a rotator with a given momentum,
restricting ourselves, for simplicity, to quadratic Hamiltonians. Aligning
the basis vectors $\boldsymbol{\xi}_{k}$ with the axes of inertia of the
rotator, we present the Hamiltonian as
\begin{align}
\hat{H}=\sum A_{k}(\hat{I}_{k})^{2}.  \label{rot.11}
\end{align}
where $A_{k}$\ are inertia momenta.
In the simplest case of a completely symmetric rotator, $A_{1}=A_{2}=A_{3}=A$,
the spectrum consists of $(2J+1)^{2}$-times degenerate multiplets
$|J\,MK\rangle $ with the energy
\begin{equation*}
E_{J}=AJ(J+1).
\end{equation*}
A simple solution takes place in the case of axial symmetry. Let
the internal axis $\boldsymbol{\xi}_{3}$ be the axis of symmetry,
$A_{1}=A_{2}=A_{\bot }$. Then, $\hat{I}_{3}=K$ is conserved; however, the
multiplet $|J\,MK\rangle $ is split in $|K|$ (of course, there remains a
$2(2J+1)$-times degeneration in $M$ and $\mathrm{sign}K$):
\begin{align*}
E_{JK} & =\langle J\,MK|A_{\bot}\left( (\hat{I}_{1})^{2}
 +(\hat{I}_{2})^{2}\right) +A_{3}(\hat{I}_{3})^{2}|J\,MK\rangle \\
& =A_{\bot}\left( J(J+1)-K^{2}\right)+A_{3}K^{2}=A_{\bot}J(J+1)+(A_{3}-A_{\bot})K^{2}.
\end{align*}

A rotator with an arbitrary relation of inertia momenta requires a more
detailed analysis \cite{Zare88}. In the basis $|J\,MK\rangle $, the
Hamiltonian (\ref{rot.11}) has non-vanishing matrix elements, $H_{KK^{\prime
}}$, with $\Delta K=0,\pm 2$, i.e., only a mixture of states with a definite
parity $K$ is admitted,
\begin{align*}
& \langle J\,MK|\hat{H}|J\,MK\rangle=\frac{1}{2} (A_{1}+A_{2})[J(J+1)-K^{2}]+A_{3}K^{2}, \\
& \langle J\,MK|\hat{H}|J\,MK\!\!+\!2\rangle
=\frac{1}{4} (A_{1}-A_{2})[(J-K)(J-K-1)(J+K+1)(J+K+2)]^{1/2},
\end{align*}
whereas the degeneration of energy levels in $\mathrm{sign}K$ is removed.

Thus, the quantum number $K$, being an eigenvalue of the right generator
$\hat{I}_{3}$ of the rotation group, plays an important role in the
description of a quantum rotator, and, correspondingly, in molecular and
nuclear spectroscopy.

For a completely symmetric rotator, not only the left transformations, but
also each of the right transformations, are symmetry transformations of the
Hamiltonian (\ref{rot.11}), the symmetry group being $SO(3)\times SO(3)$.
In the case of axial symmetry, only the right rotation, with the generator
$\hat{I}_{3}$, around the axis $\boldsymbol{\xi} _{3}$ is a symmetry of the
body, the symmetry group being $SO(3)\times SO(2)$. This symmetry
corresponds to an additive quantum number $K$. Finally, in the case of three
different momenta of inertia, the body is not symmetric, and, accordingly,
the right transformations with generators $\hat{I}_{k}$ are not its
symmetries, the symmetry group being $SO(3)$.

\textit{Consequently, whereas the symmetry with respect to the left
transformations (changes of the s.r.f.) is interpreted
as symmetries of the embedding space, in which the object is contained
(in this case, a rigid body), or as external symmetries, the symmetry with
respect to the right transformations (changes of the b.r.f) is interpreted
as symmetries of the object itself, or as internal symmetries.}

In terms of the Euler angles, the expressions for the generators
(\ref{rot.J3})--(\ref{rot.I3}) look quite complicated, as well as the
composition law. In many cases, it is more convenient to use, instead of the
Euler angles $\psi ,\theta ,\phi $, complex-valued Cayley--Klein
parameters
\begin{align}
z^{1}=\cos(\theta/2)e^{i(-\phi+\psi)/2},\qquad
z^{2}=i\sin(\theta /2)e^{i(\phi+\psi)/2},  \label{rot.12}
\end{align}
which are transformed by a spinor representation of the group $SU(2)\sim SO(3)$.
By introducing $2\times 2$ matrices, $E=\sigma ^{i}\boldsymbol{e}_{i}$
and $\Xi =\sigma ^{k}\boldsymbol{\xi}_{k}$, formula (\ref{rot.1}), that
determines the relation between s.r.f. and b.r.f., can be presented in the form
\begin{align}
\Xi = Z^{\dagger}EZ, \quad  Z^{\dagger}=Z^{-1}, \quad
Z=\left( \begin{array}{cc}
 z^1_{\;\;1} & z^1_{\;\;2} \\
 z^2_{\;\;1} & z^2_{\;\;2} \\
\end{array}\right)
=\left( \begin{array}{cc}
 z^1 & -\cc z^2 \\
 z^2 & \cc z^1 \\
\end{array}\right)\in SU(2).
\label{rot.Z}
\end{align}
Rotations of s.r.f. (\ref{rot.left}) and and b.r.f. (\ref{rot.right})
correspond to transformations in terms of unitary matrices $U$ and $\underline{U}$,
\begin{align}
Z^{\prime}=U^{\dagger}Z\underline{U},\quad U,\underline{U}\in SU(2),
\label{rot.13}
\end{align}
and, therefore, the elements of the matrix $Z$, according to (\ref{rot.Z}),
have two kinds of spinor indices: the first one being left (external), the
second one being right (internal).

The coordinates of the vector $\boldsymbol{x}=x^{i}\boldsymbol{e}_{i}$ under
the rotations of the reference frame $\{\boldsymbol{e}_{i}\}$ change as
follows:
\begin{align}
X^{\prime}=U^{\dagger}XU,\quad X=\sigma_{i}x^{i},
\label{rot.14}
\end{align}
where the matrices $U$ and $-U$ correspond to the same transformation.

Using (\ref{rot.14}) and the relation
$\overset{\ast}{U}=\sigma_{2}U\sigma_{2}$, it is easy to see that
$\sigma_{k\;\;\beta}^{\;\;\alpha}=(\sigma _{k})_{\;\;\beta}^{\alpha}$ is an
invariant tensor. A consequence of the unimodularity of the $2\times2$ matrix
$U$ is the presence of an invariant antisymmetric tensor
$\varepsilon^{\alpha\beta}=-\varepsilon^{\beta\alpha}$, $\varepsilon^{12}=\varepsilon_{21}=1$.
This allows one to lower and rise the spinor indices,
$z_{\alpha}=\varepsilon_{\alpha\beta}z^{\beta},\quad z^{\alpha}=\varepsilon^{\alpha\beta}z_{\beta}$.

In terms of the variables $z_{\;\;a}^{\alpha}$ and derivatives $\partial
_{\alpha}^{\;\;a}=\partial/\partial z_{\;\;a}^{\alpha}$, the generators take
the form
\begin{align}
\hat{J}_{k}=\frac 12(\sigma_{k})_{\;\;b}^{a}z_{\;\;a}^{\beta}\partial_{\alpha }^{\;\;a},\qquad
\hat{I}_{k}=\frac 12(\sigma_{k})_{\;\;b}^{a}z_{\alpha}^{\;\;b}\partial_{\;\;a}^{\alpha}\,,
\label{rot.15}
\end{align}

An explicit form of the states $|J\,M\,K\rangle$ is given by polynomials of
$2J$-th degree placed in the following tables
\begin{align}
J=1/2:\quad
\begin{array}{rrr}
M \backslash K& 1/2 & -1/2 \\
-1/2\quad & z^1 & \cc z^2 \\
 1/2\quad & z^2 & \cc z^1 \\
\end{array}
\qquad
J=1:\quad
\begin{array}{rccc}
M \backslash K& 1 & 0 & -1 \\
-1\quad & (z^1)^2  & z^1\cc z^2 &  (\cc z^2)^2 \\
 0\quad & z^1z^2 & z^1\cc z^1\!-\!z^2\cc z^2 &  \cc z^1\cc z^2 \\
 1\quad & (z^2)^2  & \cc z^1 z^2 & (\cc z^2)^2 \\
\end{array}
\label{rot.16}
\end{align}
The polynomial of second degree
$(-1/2)z^{\beta}_{\;\;\alphar}z_{\beta}^{\;\;\alphar} = z^1\cc z^1\!+\!z^2\cc z^2=1$,
being absent from (\ref{rot.16}), is a group invariant. A scalar product,
defined by integration with the invariant measure $d\mu (z)$ on the group $SU(2)$,
\begin{align*}
\int \cc f_1(z) f_2(z) d\mu(z), \quad
d\mu(z)= \frac{1}{8\pi^2}\delta (|z^1|^2+|z^1|^2-\!1)\, d^2z^1 d^2z^2 = \frac{1}{8\pi^2} \sin\theta d\theta d\phi d\psi,
\end{align*}
allows one to verify the orthogonality of the states (\ref{rot.16}) and
obtain the normalization coefficients.

As mentioned above, a simultaneous consideration of left and right
transformations implies an analysis of representations of the direct
product $SU(2)\times SU(2)$. The irreps of $SU(2)\times SU(2)$ are
characterized by eigenvalues of \textit{two different} Casimir operators
(the operators of squared total momentum) $\hat{J}^{2}$ and $\hat{I}^{2}$.
However, in the case under consideration $\hat{J}^{2}=\hat{I}^{2}$, and the
states are characterized only by three numbers: the total momentum $J$ and the
two projections $M$, $K$. This is a consequence of the fact that in the case
under consideration the commuting sets act in a space of functions depending
merely on three parameters. In this space, one can only construct a part of
representations of the direct product.

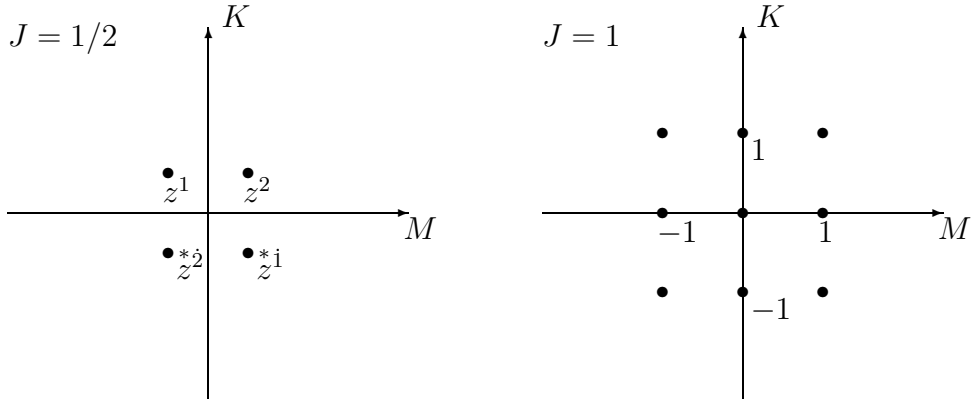
\newsavebox{\Jone}
\savebox{\Jone}(160,150)[lb]
{
\put(0,70){\vector (1,0){150}}
\put(75,0){\vector (0,1){140}}
\put(0,133){$J=1$}
\put(75,40){\circle*{4}}
\put(78,30){$-1$}
\put(75,100){\circle*{4}}
\put(78,90){$1$}
\put(45,70){\circle*{4}}
\put(43,59){$-1$}
\put(105,70){\circle*{4}}
\put(103,59){$1$}
\put(75,70){\circle*{4}}
\put(45,40){\circle*{4}}
\put(45,100){\circle*{4}}
\put(105,40){\circle*{4}}
\put(105,100){\circle*{4}}
\put(148,60){$M$}
\put(80,140){$K$}
}

\begin{picture}(360,150)
\put(0,70){\vector (1,0){150}}
\put(75,0){\vector (0,1){140}}
\put(0,133){$J=1/2$}
\put(90,55){\circle*{4}}
\put(93,45){$\cc z^{\dot 1}$}
\put(60,55){\circle*{4}}
\put(63,45){$\cc z^{\dot 2}$}
\put(90,85){\circle*{4}}
\put(88,74){$z^2$}
\put(60,85){\circle*{4}}
\put(58,74){$z^1$}
\put(148,60){$M$}
\put(80,140){$K$}
\put(200,0){\usebox{\Jone}}
\end{picture}

For the sake of clarity, the figure shows the weight diagrams of
representations with $J=1/2$ and $J=1$. For the left transformations, one
mixes the states horizontally, and for the right transformations,
vertically. In particular, at $J=1$, considering only the left or only the
right transformations (respectively, at fixed eigenvalues $\hat{I}_{3}$ and
$\hat{J}_{3}$), we obtain two different sets of three equivalent irreps
(in the general case, the number of equivalent irreps in the expansion will
be obviously equal to the dimension of this irrep). However, if one examines
both kinds of transformations at the same time, then all the nine states with
$M,K=-1,0,1$ turn out to be related by the rising and lowering operators
$\hat{J}_{\pm }$, $\hat{I}_{\pm }$. That is, the diagram of states of a rotator
with a fixed total momentum $J$ coincides with the weight diagram of the
representation $T_{J,J}$ of the direct product $SU(2)\times SU(2)$.

In the above theory of a non-relativistic rotator, the wave functions $f(z)$
depend only on its orientation. Extending the consideration to translations
implies a transition from functions $f(z)$ on the group ${\rm Spin(3)}=SU(2)$ to
functions $f(x,z)$ on $M(3)=T(3)\ctimes {\rm Spin}(3)$ -- the
group of motions of a three-dimensional Euclidean space. The position
(coordinates $x_{i}$) and orientation are then determined by two matrices
$(X,Z)$, whereas the transformations $M(3)$ -- translations and rotations --
by two matrices $(A,U)$.

For a translation and a consequent rotation of the reference frame
$\{\boldsymbol{e}_{k}\}$ by a vector $\boldsymbol{a}$, we have

\begin{equation}
X^{\prime}=U^{\dagger}(X-A)U,\quad Z^{\prime}=U^{-1}Z,\quad
A=\sigma_{i}a^{i},  \label{42}
\end{equation}
where $a^{i}$\ are the components of $\boldsymbol{a}$ in the reference
frame $\{\boldsymbol{\xi}_{k}\}$, whereas for a translation and consequent
rotation of the reference frame $\{\boldsymbol{\xi}_{k}\}$ by a vector
$\boldsymbol{b}$, we have

\begin{equation}
X^{\prime}=X+ZBZ^{\dagger},\quad Z^{\prime}=Z\underline{U},\quad B=\sigma_{i}b^{i},
\label{43}
\end{equation}
where $b^{i}$\ are the components of $\boldsymbol{b}$ in the reference
frame $\{\boldsymbol{\xi}_{k}\}$.

The orientations of b.r.f. and s.r.f. are still
related by formula (\ref{rot.Z}). However, for the generators of rotations,
instead of (\ref{rot.15}), we obtain the formulas (similar to (\ref{e2d.15}),
(\ref{e2d.15a}) in the two-dimensional case)
\begin{align}
\hat J_k = \hat L_k + \hat S_k^L = -i\varepsilon_{ijk}x_i\partial/\partial x_j
+ \frac 12(\sigma_{k})_{\;\;b}^{a} z^{\beta}_{\;\;\alphar} \partial_{\alpha}^{\;\;\alphar},
\qquad
\hat I_k = \hat S_k^R = \frac 12(\sigma_{k})_{\;\;b}^{a} z^{\;\;\betar}_\alpha \partial_{\;\;\alphar}^\alpha,
\end{align}
that is, the right generators still contain only the intrinsic momentum in
b.r.f., whereas the left ones are the sum of the orbital $\hat{L}_{k}$ and
intrinsic $\hat{S}_{k}^{L}$ momenta.

\section{Description of orientable objects and changes \newline
of reference frames}

The position of a point-like object in a $d$-dimensional
Euclidean space is describe by space coordinates, $x^{k}$, $k=1,\ldots,d$
(respectively, by space-time coordinates $x^{\mu}$, $\mu=0,1,\ldots,d-1$,
in a pseudo-Euclidean space).

For an orientable object, nevertheless, be it a rigid body or a spinning
elementary particle, such a treatment is obviously incomplete.

Whereas for a description of non-orientable objects it is sufficient to use
\textit{one} s.r.f., for a description of orientable objects it is convenient
to use \textit{two} orthonormalized reference frames: a s.r.f. $\{e_{i}\}$ and
a b.r.f. $\{\xi_{k}\}$ reference frame,
\begin{align}
\xi_{k}=v_{\;k}^{i}e_{i}.
\label{orient.1}
\end{align}
For Euclidean spaces, $(e_{i},e_{j})=\delta_{ij}$, and the matrix $V=\Vert
v_{\;k}^{i}\Vert$, composed of the coefficients of a re-decomposition of the
bases (s.r.f. and b.r.f.), is orthogonal, $V^{-1}=V^{T}$. For pseudo-Euclidean
spaces (and, in particular, the four-dimensional Minkowski space) the matrix
$V$ is pseudo-orthogonal, $V^{-1}=\eta V^{T}\eta$, $\eta=\mathrm{diag}(1,-1,\dots,-1)$.

That is, besides $d$ spatial coordinates $x^{\mu}$, a description of an
orientable object requires to use $d(d-1)/2$ parameters that determine a
$d\times d$ (pseudo)orthogonal matrix $V$. Thus, the complete set of its
coordinates is $(x^{i},v_{\;\,k}^{i}).$

Let us consider the transformation law for $(x^{i},v_{\;\,k}^{i})$ at the
changes of s.r.f. and b.r.f. Consider first rotations. In the matrix notation
(\ref{orient.1}) reads $\xi=eV$. A rotation of the s.r.f. $e^{\prime}=e\Lambda,$
where $\Lambda$ is the matrix of rotations, yields $\xi =eV=e^{\prime}\Lambda^{-1}V$,
whence $V^{\prime}=\Lambda^{-1}V$. A rotation of the b.r.f.
$\xi^{\prime}=\xi\underline{\Lambda}$ yields $\xi=eV=\xi'\underline{\Lambda}^{-1}$,
whence $V^{\prime}=V\underline{\Lambda}$. Both for
rotations and translations, simple calculations lead to the following results.

For rotations and translations, simple calculations lead to the following
results.

A rotation of the s.r.f. with a consequent translation:
\begin{align}
x^{\prime}=\Lambda^{-1}(x-a),\quad V^{\prime}=\Lambda^{-1}V, \label{orient.l1}
\end{align}
where the column $a$ is composed of the coordinates $a_{\mu }$ of the
translation vector in the s.r.f..
A translation of the b.r.f. with a consequent rotation:
\begin{align}
x^{\prime}=x+Vb,\quad V^{\prime}=V\underline{\Lambda},  \label{orient.b1}
\end{align}
where the column $b$ is composed of the coordinates of the translation
vector in the b.r.f.; $Vb$ are the coordinates of the same vector in the s.r.f..
It is easy to see that rotations of the s.r.f. do not affect the coordinates
of the body $x$, but only its orientation $V$.

Let us now present these transformations in terms of the group of motions.
The group of motions of a Euclidean space is a group of transformations that
preserves the distance
\begin{align}
r^{2}=\delta_{ik}(x^{i}-y^{i})(x^{k}-y^{k})  \label{orient.2}
\end{align}
between two points. The group of motions of a pseudo-Euclidean space preserves
the interval
\begin{align}
s^{2}=\eta_{\mu\nu}(x^{\mu}-y^{\mu})(x^{\nu}-y^{\nu}),  \label{orient.3}
\end{align}
where $\eta =\mathrm{diag}(1,-1,\dots ,-1)$ is the metric tensor of the
Minkowski space.

The transformations consist of translations and rotations (for a
pseudo-Euclidean space also of boosts (hyperbolic rotations)). If $x$ are
the coordinates of a point in a basis $\{e_{i}\}$, then, as a result of a
rotation and translation, we obtain a point with the coordinates
\begin{align}
x'=\Lambda x+a.  \label{orient.4}
\end{align}
Each element $g$ of the group of motions corresponds to a pair of matrices,
$g=(a,\Lambda)$, where $a$ is a column of the elements $a^{i}$ or $a^{\mu}$,
corresponding to translations; $\Lambda$ is a matrix of $SO(d)$ or $SO(d-1,1) $.

Those transformations (\ref{orient.4}) that can be continuously connected with
the identity form a Lie group called the proper Poincar\'{e} group
$M_{0}(d-1,1)$. The corresponding homogenous transformations ($a=0$) form the
proper Lorentz group $SO_{0}(d-1,1)$.
In the Euclidean space, we have $M_{0}(d)$ and $SO_{0}(d)$, respectively. The
law of composition and the inverse element of the Poincar\'{e} group have
the form
\begin{align}
(a_{2},\Lambda_{2})(a_{1},\Lambda_{1})=(a_{2}+\Lambda_{2}a_{1},\Lambda _{2}\Lambda_{1}), \qquad
g^{-1}=(-\Lambda^{-1}a,\Lambda^{-1}),
\label{orient.5}
\end{align}
whence it follows that the groups $M_{0}(d-1,1)$ and $M_{0}(d)$ are
semi-direct products:
\begin{equation*}
M_{0}(d-1,1)=T(d)\ctimes SO_{0}(d-1,1),\qquad
M_{0}(D)=T(d)\ctimes SO_{0}(d),
\end{equation*}
where $T(d)$ is the group of $d$-dimensional translations.

Note that an element $g$ of the group $M_{0}(d)$ or $M_{0}(d-1,1)$ can also
be associated with one $(d+1)\times (d+1)$ matrix, whereas the
transformation (\ref{orient.4}) takes the form
\begin{equation*}
g\longleftrightarrow
\left( \begin{array}{cc} \Lambda & a \\ 0 & 1 \end{array} \right) , \qquad
\left( \begin{array}{c} x^{\prime} \\ 1 \end{array} \right)
=\left(\begin{array}{cc} \Lambda & a \\ 0 & 1 \end{array} \right)
\left( \begin{array}{c} x \\ 1 \end{array} \right) .
\end{equation*}

A pair $(x,V)$ uniquely determines the position and orientation of the
b.r.f. with respect to the s.r.f..
That is, the manifold of the Poincar\'e group (as observed in
\cite{Lurca64,Tolle78}) is isomorphic to the space of all b.r.f.
This pair can be associated with an element of the group $q$,
$q\leftrightarrow(x,V)$. It is easy to see that \textit{a change of
the s.r.f.} (\ref{orient.l1}) \textit{corresponds to
left-multiplication} by group elements $g^{-1}$:
\begin{align}
q'= g^{-1}q\longleftrightarrow(x',V')
  = (a,\Lambda )^{-1}(x,V)=(\Lambda^{-1}(x-a),\ \Lambda^{-1}V),  \label{orient.l2}
\end{align}
whereas \textit{a change of the b.r.f.} (\ref{orient.b1})
\textit{corresponds to right-multiplication} by group elements $h$:
\begin{align}
q' = qh\longleftrightarrow(x',V')=(x,V)(b,\underline {\Lambda})
   = (x+Vb,\;V\underline{\Lambda}).  \label{orient.b2}
\end{align}
In the matrix form, the general transformation $q^{\prime}=g^{-1}qh$ is
given by
\begin{align}
\left(
\begin{array}{cc}
V^{\prime} & x^{\prime} \\
0 & 1
\end{array}
\right) =\left(
\begin{array}{cc}
\Lambda & a \\
0 & 1
\end{array}
\right) ^{-1}\left(
\begin{array}{cc}
V & x \\
0 & 1
\end{array}
\right) \left(
\begin{array}{cc}
\underline{\Lambda} & b \\
0 & 1
\end{array}
\right) .  \label{orient.6}
\end{align}

As to the choice of variables that describe the orientation of an object, one
can mention the following:

As such variables, one can select $d(d-1)/2$ independent variables (angles);
however, the law of composition presented in terms of these angles is
sufficiently involved even in the case of $d=3$, not to mention the higher
dimensions.

One can select $d^{2}$ real-valued parameters, being the elements
$v^{i}_{\;k}$ of a (pseudo)orthogonal matrix $V$, which is the matrix of the
vector representation of the group $SO_{0}(d)$ or $SO_{0}(d-1,1)$.

Having in mind the different transformation laws (\ref{orient.l2}) and
(\ref{orient.b2}), we are going to denote the elements of the matrix $V$ as
$v_{\;\;\underline{n}}^{\mu }$, with the Greek indices $\mu ,\nu ,\ldots $
for the left transformations of the Poincar\'{e} group and with the
underlined Latin indices $\mur,\nur,\ldots $ for the
right transformations. Note that in 3+1 dimensions $v_{\;\;\nur}^{\mu }$
are given by tetrads, i.e., objects that transform as vectors (in
the first index $\mu $) under the change of a laboratory (space-fixed)
reference frame and as vectors (in the second index
$\nur$) under the change of a local (body-fixed) reference frame.

On the other hand, one can select elements of the spinor representation
for spaces of three and four dimensions: these are the elements of a
complex-valued $2\times 2$ matrix $Z=\Vert z_{\;\;\betar}^{\alpha }\Vert $.
The composition law in this case has an especially simple form;
besides, such a parameterization allows one to describe half-integer spins.
In other words, having at one's disposal $z_{\;\;\betar}^{\alpha }$,
one can obtain $v_{\;\;\nur}^{\mu }$, whereas the inverse
operation is two-fold. It is this parameterization that will be used in what
follows. We underline ``right-hand'' indices in order to avoid
misunderstanding, since we shall examine quantities with fixed values of
indices (for instance, spinors $z_{\;\;\underline{1}}^{\alpha }$ and
$z_{\;\;\underline{2}}^{\alpha }$).

\section{Parameterization of the Poincar\'{e} group}

First we recall that there exists a well-known one-to-one correspondence
between the 4-vectors $x$ and $2\times 2$ Hermitian matrices $X$ {\footnote{%
We use two sets of $2\times 2$ matrices $\sigma _{\mu }=(\sigma _{0},\sigma
_{k})$ and $\bar{\sigma}_{\mu }=(\sigma _{0},-\sigma _{k})$,
 \begin{equation*} 
 \sigma_0=\left(\begin{array}{cc} 1 & 0  \\ 0 & 1 \end{array}\right),\quad
 \sigma_1=\left(\begin{array}{cc} 0 & 1  \\ 1 & 0 \end{array}\right),\quad
 \sigma_2=\left(\begin{array}{cc} 0 & -i \\ i & 0 \end{array}\right),\quad
 \sigma_3=\left(\begin{array}{cc} 1 & 0 \\ 0 & -1 \end{array}\right).
 \end{equation*}
},}
\begin{equation} \label{par.2}
 X=x^\mu\sigma_\mu = \left(
 \begin{array}{cc} x^0+x^3 & x^1-ix^2 \\ x^1+ix^2 & x^0-x^3 \end{array}\right), \quad
 \det X=x_\mu x^\mu ,\quad x^\mu =\frac 12\mathop{\rm Tr}(X\bar\sigma^\mu).
\end{equation}
One ought to say that this correspondence plays an important role in twistor theory
\cite{Penro68,PenMa72,Wells79}.
If $x$ is subject to the transformation
\begin{equation*}  
x'=gx, \qquad x'^\nu=\Lambda^\nu_{\;\;\mu} x^\mu + a^\nu ,
\end{equation*}
then $X$ transforms (see, for example, \cite{Penro68,Vilen68t,BarRa77}) as follows:
\begin{equation}  \label{par.4}
X'=gX=UXU^\dagger +A,
\end{equation}
where $A=a^{\mu}\sigma_{\mu}$, and the complex matrices $U$ obey the conditions
\begin{equation}  \label{ref.5}
\sigma_\nu\Lambda^\nu_{\;\;\mu}=U\sigma_\mu U^\dagger.
\end{equation}

The representation of the Poincar\'e transformations in the form (\ref{par.4})
is closely related to a representation of finite rotations in $\mathbb{R}^{d}$
in terms of the Clifford algebra. In higher dimensions the
transformation law has the same form, where $A$ is a vector element and $U$
corresponds to an invertible element (spinor element) of the Clifford
algebra \cite{BenTu88}.

Using (\ref{ref.5}), we get $\det U=e^{i\phi }$. Matrices $U$ that differ
only by a phase factor correspond to the same $\Lambda $, and we can fix
this arbitrariness by imposing the condition $\det U=1$.
However, even after this two matrices ($U$,$-U$) correspond to one $\Lambda$.
Considering both $U$ and $-U$ as representatives for $\Lambda $, we, in fact, go over
from $SO_{0}(3,1)$ to its double covering group $\mathrm{Spin}(3,1)=SL(2,C)$,
\begin{equation}
U=\left(\begin{array}{cc} u^1_{\;1} & u^1_{\;2} \\ u^2_{\;1} & u^2_{\;2}
 \end{array}\right) \in SL(2,C) ,
 \quad u_{\;1}^1 u_{\;2}^2-u_{\;1}^2 u_{\;2}^1=1,
 \label{ref.6}
\end{equation}
and from $M_{0}(3,1)=T(4)\ctimes \mathrm{SO}(3,1)$ to
\begin{equation*}
M(3,1)=T(4)\ctimes\mathrm{Spin}(3,1).
\end{equation*}

As is known, this allows one to avoid double-valued representations for
half-integer spins. Thus, there exists a one-to-one correspondence between
elements $g$ of $M(3,1)$ and two $2\times 2$ matrices,
$g\leftrightarrow (A,U)$. The first one, $A$, corresponds to translations and
the second one, $U$, corresponds to rotations. Eq. (\ref{par.4}) describes
the action of $M(3,1)$ in the Minkowski space (the latter is the coset space
$M(3,1)/\mathrm{Spin}(3,1)$).

As a consequence of (\ref{par.4}), one can obtain the composition law and
the inverse element:
\begin{equation}
(A_{2},U_{2})(A_{1},U_{1})=(U_{2}A_{1}U_{2}^{\dagger}+A_{2},\;
U_{2}U_{1})\;,\quad g^{-1}=(-U^{-1}A(U^{-1})^{\dagger},\;U^{-1})\;.
\label{par.7}
\end{equation}
The matrices $U$ satisfy the identity
\begin{equation}
\sigma _{2}U\sigma _{2}=(U^{T})^{-1}.  \label{par.8}
\end{equation}

An equivalent picture arises in terms of the matrices $\overline{X}=x^{\mu } \bar{\sigma}_{\mu }$.
Using the relation $\overline{X}=\sigma _{2}X^{T}\sigma _{2}$,
the transformation law for $X$ (\ref{par.4}), and identity (\ref{par.8}), one can get
\begin{equation}
\overline{X}^{\prime}=(U^\dagger)^{-1}\overline{X}U^{-1}+\overline{A},
\label{xbar}
\end{equation}
Thus, $\overline{X}$ are transformed by means of the elements
$(\overline{A},(U^{\dagger })^{-1})$. The relation
$(A,U)\rightarrow (\overline{A},(U^{\dagger })^{-1})$ determines an automorphism
of the group $M(3,1)$. In the Euclidean case, the matrices $U$ are unitary, and
the latter relation is reduced to $(A,U)\rightarrow (-A,U)$.

\section{Regular representation and spin-coordinate space}

In the above chosen representation, the position and the translation parameters
are given by a Hermitian $2\times 2$ matrix $X$, whereas the orientation and
rotations are given by a complex-valued $2\times 2$ matrix, which we denote
as $Z$, $Z\in SL(2,C)$. Using the $2\times 2$ matrices $E=\sigma ^{\mu }{e}_{\mu }$
and $\Xi =\sigma ^\nur{\xi }_\nur$, we can rewrite formula (\ref{orient.1}), which
determines the mutual orientation of the laboratory and body-fixed reference frames,
in the form
\begin{equation}
\Xi =Z^{\dagger }EZ.  \label{reg.0}
\end{equation}
Making a comparison of (\ref{orient.1}) and (\ref{reg.0}), we obtain
$\sigma^{\nur}v_{\;\;\nur}^{\mu }{e}_{\mu }=Z^{\dagger}\sigma ^{\mu }{e}_{\mu }Z$,
or, in terms of the components,
\begin{equation*}
(\sigma ^\nur)_{\dot\betar\underline{a}}v_{\;\;\underline{n}}^{\mu }=
\cc z_{\;\;\dot\betar}^{\dot{\beta}}(\sigma^{\mu })_{\dot{\beta}\alpha }
z_{\;\;\underline{a}}^{\alpha }.
\end{equation*}
Multiplying this by
$(\bar{\sigma}^\mur)^{\alphar\dot{\betar}}$ and using the identity
$\mathop{{\rm Tr}}\sigma ^{\mu }\bar{\sigma}^{\nu }
=(\sigma ^{\mu })_{\alpha \dot{\beta}}(\bar{\sigma}^{\nu })^{\dot{\beta}\alpha }=2\eta ^{\mu \nu }$,
we find
\begin{equation}
v_{\;\;\underline{n}}^{\mu }=\frac{1}{2}(\sigma ^{\mu })_{\dot{\beta}\alpha}
(\bar{\sigma}_\nur)^{\underline{a}\dot\betar}z_{\;\;\underline{a}}^{\alpha }
\cc z_{\;\;\dot\betar}^{\dot{\beta}},
\label{reg.v0}
\end{equation}
where $v_{\;\;\underline{n}}^{\mu }$ are elements of pseudoorthogonal matrix
$V\in SO(3,1)$, $V^{-1}=\eta V^{T}\eta $, and obey the orthogonality conditions
\begin{equation}
v_{\;\;\nur}^{\mu }v_{\mu }^{\;\;\mur}
=\delta _{\nur}^\mur,\quad v_{\;\;\nur}^{\mu }v_{\nu }^{\;\;\nur}
=\delta _{\nu }^{\mu }.  \label{reg.v}
\end{equation}

Thus, the pair $(X,Z)$ uniquely identifies the position and orientation of
the b.r.f. with respect to the s.r.f.; besides, a change of the s.r.f. corresponds to
the left-multiplication by $(A,U)^{-1}$, while a change of the b.r.f.
corresponds to the right-multiplication by
$(\underline{A},\underline{U})$:
\begin{equation}
(X',Z')=(A,U)^{-1}(X,Z)(\underline{A},\underline{U})
=(U^{-1}(X-A)(U^{\dagger })^{-1}+Z\underline{A}Z^{\dagger },\,U^{-1}Z\underline{U}).
\label{reg.1}
\end{equation}

Let us now examine functions of the coordinates and orientation, i.e.,
functions defined on the Poincar\'{e} group,\ $f(q)$, $q\in M(3,1)$. The
action of the group $M(3,1)\times M(3,1)$ in the space of functions $f(q)$
is given by
\begin{align}
&\mathbb{T}(g,h)f(q)=f'(q)=f(g^{-1}qh), \label{reg.2}
\\ \label{reg.3}
&q\leftrightarrow (X,Z),\quad  g\leftrightarrow (A,U),\quad h\leftrightarrow (\underline{A},\underline{U}).
\end{align}
As a consequence of\ (\ref{reg.2}), we have
\begin{equation}
f'(q')=f(q),\quad q'=gqh^{-1}.  \label{reg.3a}
\end{equation}
In view of the relations
\begin{equation}
X=x^{\mu }\sigma _{\mu },\quad Z=\left(
\begin{array}{cc}
z_{\;\,\underline{1}}^{1} & z_{\;\,\underline{2}}^{1} \\
z_{\;\,\underline{1}}^{2} & z_{\;\,\underline{2}}^{2}
\end{array}
\right) \in SL(2,C),  \label{reg.6}
\end{equation}
the mapping $q\leftrightarrow (X,Z)$ leads to the correspondence
\begin{align}
& q\leftrightarrow (x,z),\quad
\hbox{where}\quad x=(x^{\mu }),\;z=(z_{\;\,{\betar}}^{\alpha }),
\label{reg.7} \\
& \mu =0,1,2,3,\quad
\alpha ,b=1,2,\;\quad
z_{\;\,\underline{1}}^{1}z_{\;\,\underline{2}}^{2}-z_{\;\,\underline{1}}^{2}z_{\;\,\underline{2}}^{1}=1,
\nonumber
\end{align}
and relation\ (\ref{reg.3a}) takes the form
\begin{equation}
f'(x',z')=f(x,z),\quad (x',z^{\prime})\leftrightarrow q'=gqh^{-1}.  \label{reg.3b}
\end{equation}

The action of the left GRR $T_{L}(g)$ in the space of functions $f(q)$ on
the group
\begin{equation}
T_{L}(g)f(q)=f'(q)=f(g^{-1}q),  \label{reg.4}
\end{equation}
is related to the change of the s.r.f.; see (\ref{orient.l2});
and, correspondingly,
\begin{equation}
f'(q')=f(q),\quad q'=gq.  \label{reg.5}
\end{equation}
On the other hand, we have the correspondence $q'\leftrightarrow
(x',z')$,
\begin{align*}
& q'=gq\leftrightarrow (X',Z')=(A,U)(X,Z)=(UXU^{+}+A,UZ)\leftrightarrow (x',z'), \\
& X'=UXU^{+}+A\quad \Longrightarrow \quad
 x^{\prime \mu }=\Lambda_{\;\,\nu }^{\mu }x^{\nu }+a^{\mu },\quad  \\
& Z'=UZ\quad \Longrightarrow \quad z_{\;\;\;\betar}^{\prime \alpha }
 =U_{\;\;\beta }^{\alpha }z_{\;\,\betar}^{\beta},\quad U=(U_{\;\;\beta }^{\alpha }),
\end{align*}
Then, relation (\ref{reg.5}) takes the form
\begin{align}
&f'(x',z')=f(x,z),  \label{reg.8}
\\ \label{reg.9}
&x^{\prime\mu}=\Lambda_{\;\,\nu}^\mu x^\nu+a^\mu, \quad
 \Lambda\in SO_0(3,1)\leftarrow U\in SL(2,C),
\\ \label{reg.10}
& z^{\prime\alpha}_{\;\;\;\betar}=U^{\alpha}_{\;\;\beta}z^{\beta}_{\;\;\betar}. \quad
\end{align}

Relations (\ref{reg.8})--(\ref{reg.10}) admit a remarkable interpretation.
We may treat $x$ and $x'$ in these relations as position coordinates in the Minkowski
space $M(3,1)/SL(2,C)$ (in different Lorentz reference frames) related by the proper
Poincar\'{e} transformations, and the sets $z$ and $z'$ (coordinates on $SL(2,C)$)
may be treated as spin coordinates in such frames. Carrying two-dimensional spinor
representation of the Lorentz group, the variables
$z_{\;\;\underline{1}}^{\alpha }$ and $z_{\;\;\underline{2}}^{\alpha }$
are invariant under translations as one can expect for spin degrees of freedom.

Therefore, we may treat the sets $(x,z)$ as points in the position-spin
space with the transformation law (\ref{reg.9}), (\ref{reg.10}) under the
change from one Lorentz reference frame to another. In this case, equations
(\ref{reg.8})--(\ref{reg.10}) present the transformation law for scalar
functions in the position-spin space.

On the other hand, as has been demonstrated, the sets $(x,z)$ are in
one-to-one correspondence to group $M(3,1)$ elements. Thus, the
functions (\ref{reg.3b}) are still functions on this group. For this reason,
we often call them scalar functions on the group as well, recalling that the
term ``scalar'' originates from the above interpretation.

Generally speaking, the functions $f(x,z)$ are not analytical functions of
the complex variables $z_{\;\;\betar}^{\alpha }$, i.e., they
depend on both $z_{\;\;\betar}^{\alpha }$ and the complex-conjugate
$\cc{z}_{\;\;\dot{\beta}}^{\dot{\alpha}}$. Correspondingly, we shall later
regard the functions $f(x,z)$ as functions of the variables
$x^{\mu},\,z_{\;\;\betar}^{\alpha },\,\cc{z}_{\;\;\dot{\beta}}^{\dot{\alpha}}$.

Consider now the right GRR $T_{R}(h)$. This representation is defined in the
space of functions $f(q)$, $q\in M(3,1)$ as follows:
\begin{equation}
T_{R}(h)f(q)=f'(q)=f(qh),\quad   \label{reg.11}
\end{equation}
The action of a right GRR corresponds to a change of the b.r.f.; see (\ref{orient.b2}).
As a consequence of relation (\ref{reg.11}), one can write
\begin{equation}
\label{reg.12}
f'(q')=f(q),\quad
q'=qh^{-1} \leftrightarrow (X',Z')=(X-Z\underline{U}^{-1}\underline{A}(\underline{U}^{-1})^{\dagger}Z^{\dagger },Z\underline{U}^{-1}),
\end{equation}
whence
\begin{align}
&f'(x',z')=f(x,z),  \nonumber
\\
&x^{\prime\mur}=x^{\mur}-(\underline{\Lambda}^{-1})_{\;\;\:\nur }^\mur a^\nur
\quad {\mbox {or}} \quad
x^{\prime\mu} =x^\mu - v_{\;\;\mur }^\mu(\underline{\Lambda}^{-1})_{\;\;\,\nur }^\mur a^\nur ,\quad
\underline{\Lambda}\in SO_0(3,1), 
\label{reg.13}
\\
& z^{\prime\alpha}_{\;\;\;\betar}=z^{\alpha}_{\;\;\alphar}(\underline{U}^{-1})^{\alphar}_{\;\;\betar}.
\label{reg.14}
\end{align}
where\ $v_{\;\;\mur}^{\mu }$, which determine the orientation of the b.r.f.,
are expressed in terms of $z$; see (\ref{reg.v0}).

These transformations are essentially different from the Lorentz
transformations (\ref{reg.9}), (\ref{reg.10}) in the extended
spin-coordinate space. For the parameters of right translations
$a^{\underline{n}}=0$, according to (\ref{reg.13}), we have $x^{\prime \mu }=x^{\mu }$,
i.e., the right rotations lead only to a change of orientation,
and, as distinct from the left rotations, do not affect the space-time
coordinates $x^{\mu }$. On the other hand, the right translations (\ref{reg.13}),
as distinct from the left ones (\ref{reg.9}), create a
``mixture'' of the space coordinates $x$ with the spin (orientation)
coordinates $z$.

Besides, it is easy to see that, whereas the left transformations of the
group $M(3,1)$ never affect the interval,
\begin{align}
& g^{-1}X-g^{-1}Y=X'-Y'=U^{-1}(X-Y)(U^{-1})^{\dagger },
\notag  \\
& s^{\prime 2}=\det (X'-Y')=\det (X-Y)=s^{2};
 \label{reg.15}
\end{align}
the right transformations at $A\neq 0$ do not affect the interval only on
condition that $Z_{1}=Z_{2}$, which corresponds to an equal orientation of
b.r.f.:
\begin{align}
& Xh-Yh=X''-Y''=X-Y+(Z_{1}-Z_{2})A(Z_{1}^{\dagger }-Z_{2}^{\dagger }),  \notag
\\
& s^{\prime \prime 2}=\det [X-Y+(Z_{1}-Z_{2})A(Z_{1}^{\dagger }-Z_{2}^{\dagger })].
\label{reg.16}
\end{align}

The generators of the left GRR and of the right GRR are applied for a
classification of scalar functions on the Poincar\'e group, because they
enter the maximal set of commuting operators on the group.

Let us consider the space of functions on the Poincar\'e group.
An invariant measure on the group has the form
$d\mu(x,z)=d^4x d\mu(z)$, where $d\mu(z)=(i/2)^3d^2 z_{\;\,\underline{1}}^{1}
d^2 z_{\;\,\underline{1}}^{2} d^2 z_{\;\,\underline{2}}^{1} |z_{\;\,\underline{1}}^{1}|^{-2}$
is the measure on the Lorentz group \cite{GelGrV66}.
If a GRR acts in the space of all functions on the group $G$, then a regular
representation acts in the space of functions $L^{2}(G,\mu )$, such that the
norm
\begin{equation}
\int \cc f(g)f(g)d\mu (g)  \label{reg.17}
\end{equation}
is finite, where $d\mu(g)$ is an invariant measure on the group
\cite{VilKl91,ZhelSc83}.
The regular representation is unitary, as it follows
from (\ref{reg.17}), as well as from the invariance of the measure\footnote{A
decomposition of the regular representation does not include unitary irreps of
the auxiliary series, characterized by the nonlocal scalar product
$\int\cc{f}(\tilde{g}_{1})f(\tilde{g}_{2})I(\tilde{g}_{1},\tilde{g}_{2})
d\mu(\tilde{g}_{1})d\mu(\tilde{g}_{2})$, where the kernel $I(\tilde{g}_{1},\tilde{g}_{2})$
is invariant with respect to group transformations,
$\tilde{g}_{k}\in G/H$, $H\subset G$. However, such representations have not
been applied in physics so far.}.

On the other hand, polynomials in $z$ and $\cc z$ carry finite-dimensional
nonunitary representations of the Lorentz group and therefore integral (\ref{reg.17})
diverges in this case.
Thus, we use different spaces of functions on the Poincar\'e group:
$L^{2}(G,\mu )$ for unitary representations of principal series (corresponding to
infinite-component fields in Majorana type equations)
and a space of polynomials in $z,\cc z$ for finite-component representations
(corresponding to spin-tensor fields) of the Lorentz group.

\section{A field on the Poincar\'{e} group and spin-tensor fields}

We shall now discuss a relation between the description of orientable objects, in
particular, higher spins, in terms of scalar functions $f(x,z)$ on the
Poincar\'{e} group, with the standard description in terms of multi-component fields.

Spin-tensor fields that describe particles of different spins \textit{are defined
by a transformation law corresponding to a change of s.r.f..}
These fields are related to multi-component functions on the Minkowski space
(i.e., functions depending not only on $x$ but also on a certain discrete parameter).

The relation $f'(q')=f(q)$, $q'=gq$, connected with the left GRR (\ref{reg.4}),
also determines a law of field transformation with the change of s.r.f.; however, this
transformations act in the extended (spin-coordinate) space.
Scalar functions on the Poincar\'{e} group\ $f(q)$, $q=(x,z)$, depend
not only on $x$ but also on the set of the variables $z$:
\begin{align}
f'(x',z')=f(x,z), \qquad
x'=gx=\Lambda x+a\leftrightarrow UXU^{\dagger }+A,\quad
z'=gz\leftrightarrow UZ.  \label{tens.1}
\end{align}
In contrast to a scalar field in the Minkowski space, this field is reducible
not only with respect to mass, but with respect to spin as well.
It is well known \cite{BarRa77,ZhelSc83,VilKl91} that any irrep of the group $G$ is
contained (up to an equivalence) in the decomposition of the left (or right) GRR.
Thus, the task of classification of $M(3,1)$ irreps reduces to the task of
classification of scalar functions (\ref{tens.1}).

In this section, we examine only left transformations; besides, for the sake
of brevity, we omit the second (``right'') index of $z_{\;\;\betar}^{\alpha }$,
because the elements of the first and second columns $z_{\;\;\underline{1}}^{\alpha }$,
$z_{\;\;\underline{2}}^{\alpha }$ of the matrix (\ref{reg.6}) transform under the
action of the left GRR of $M(3,1)$ in the same way.

According to (\ref{tens.1}) and (\ref{xbar}), one can write the transformation law
of $x^{\mu }$, $z_{\alpha }$, $\cc z_{\dot{\alpha}}$ in component-wise form
\begin{align}
& x^{\prime \nu }\sigma _{\nu \alpha {\dot{\alpha}}}
 =U_{\alpha }^{\;\;\beta}x^{\mu }\sigma _{\mu \beta {\dot{\beta}}}\cc U_{\;\;\dot{\alpha}}^{\dot{\beta}}
 +a^{\mu }\sigma _{\mu \alpha {\dot{\alpha}}},\quad
 x^{\prime \nu }\bar{\sigma}_{\nu }^{\;\;{\dot{\alpha}}\alpha }
 =(\cc U^{-1})_{\;\;\dot{\beta}}^{\dot{\alpha}}x^{\mu }\bar{\sigma}_{\mu }^{\;\;{\dot{\beta}}\beta}(U^{-1})_{\beta }^{\;\;\alpha }
 +a^{\mu }\bar{\sigma}_{\mu }^{\;\;{\dot{\alpha}}\alpha },\qquad
\label{tens.2} \\
& z_{\alpha }'=U_{\alpha }^{\;\;\beta }z_{\beta },\quad \cc z_{{\dot{\alpha}}}'
 =\cc U_{{\dot{\alpha}}}^{\;\;{\dot{\beta}}}\cc z_{{\dot{\beta}}},\quad
 z^{\prime \alpha }=(U^{-1})_{\;\;\beta }^{\alpha}z^{\beta },\quad
 \cc z^{\prime {\dot{\alpha}}}=(\cc U^{-1})_{\;\;{\dot{\beta}}}^{{\dot{\alpha}}}\cc z^{{\dot{\beta}}}.
\label{tens.2a}
\end{align}
It is easy to see from (\ref{tens.2}) that the tensors
\begin{equation}
\sigma _{\mu \alpha {\dot{\alpha}}}=(\sigma _{\mu })_{\alpha {\dot{\alpha}}},\quad
\bar{\sigma}_{\mu }^{\;\;{\dot{\alpha}}\alpha }=(\bar{\sigma}_{\mu})^{{\dot{\alpha}}\alpha },
\label{tens.3}
\end{equation}
are invariant. These tensors are usually applied to convert the vector
indices into spinor ones and vice versa, or to construct vector from
two spinors of different types:
\begin{equation}
x_{\alpha {\dot{\alpha}}}=(X)_{\alpha {\dot{\alpha}}}=\sigma _{\mu \alpha {\dot{\alpha}}}x^{\mu },\quad
x^{\mu }=\frac{1}{2}\bar{\sigma}^{\mu {\dot{\alpha}}\alpha }x_{\alpha {\dot{\alpha}}},\quad
q^{\mu }=\frac{1}{2}\bar{\sigma}^{\mu {\dot{\alpha}}\alpha }z_{\alpha }\cc z_{\dot{\alpha}}.
\label{tens.4}
\end{equation}
In consequence of the unimodularity of $2\times 2$ matrices $U$ there
exist invariant antisymmetric tensors
$\varepsilon ^{\alpha \beta}=-\varepsilon ^{\beta \alpha }$,
$\varepsilon ^{{\dot{\alpha}}{\dot{\beta}}}=-\varepsilon ^{{\dot{\beta}}{\dot{\alpha}}}$,
$\varepsilon ^{12}=\varepsilon ^{{\dot{1}}{\dot{2}}}=1$,
$\varepsilon _{12}=\varepsilon _{{\dot{1}}{\dot{2}}}=-1$.
Now spinor indices are lowered and raised according to the rules
\begin{equation}
z_{\alpha }=\varepsilon _{\alpha \beta }z^{\beta },\quad
z^{\alpha }=\varepsilon ^{\alpha \beta }z_{\beta },
\label{tens.5}
\end{equation}
and in particular one can get
$\bar{\sigma}_{\mu {\dot{\alpha}}\alpha}\equiv
\varepsilon _{\dot{\alpha}\dot{\beta}}\varepsilon _{\alpha \beta }\bar{\sigma}_{\mu }^{\;\;{\dot{\beta}}\beta }
=\sigma _{\mu \alpha {\dot{\alpha}}}$.
Below we will also use the notations
\begin{equation}
\partial _{\alpha }={\partial }/{\partial z^{\alpha }},\quad \partial ^{\dot{\alpha}}
={\partial }/{\partial \cc z_{\dot{\alpha}}},\quad \partial ^{\alpha}
=\varepsilon ^{\alpha \beta }\partial _{\beta }=-{\partial }/{\partial z_{\alpha }},\quad
\partial _{\dot{\alpha}}=\varepsilon ^{{\dot{\alpha}}{\dot{\beta}}}\partial _{\dot{\beta}}
=-{\partial }/{\partial \cc z^{\dot{\alpha}}}.
\label{tens.6}
\end{equation}

In the framework of the present approach, \textit{a standard spin
description in terms of multicomponent functions arises under the separation
of space and spin variables}.

Since $z$ is invariant under translations, any function $\phi(z)$ carries a
representation of the Lorentz group. Let a function $f(x,z)$ allow the
representation
\begin{equation}
f(x,z)=\phi^{n}(z)\psi_{n}(x),  \label{tens.7}
\end{equation}
where $\phi^{n}(z)$ form a basis in the representation space of the Lorentz
group. The latter means that one can decompose the functions $\phi^{n}(z')$
of the transformed argument $z'=gz$ in terms of the functions $\phi^{n}(z)$:
\begin{equation}
\phi^{n}(z')=\phi ^{l}(z)L_{l}^{\;\;n}(U).  \label{tens.8}
\end{equation}
An action of the Poincar\'{e} group on a line $\phi (z)$ composed of
$\phi^{n}(z)$ is reduced to the multiplication by a matrix $L(U)$, where
$U\in \mathrm{Spin}(3,1)$, $\phi (z')=\phi (z)L(U)$.

As one compares the decompositions of the function $f'(x',z')=f(x,z)$ over
the transformed basis $\phi (z')$ and over the initial basis $\phi (z)$,
\begin{equation*}
f'(x',z')=\phi (z')\psi'(x')=\phi (z)L(U)\psi '(x')=\phi (z)\psi (x),
\end{equation*}
where $\psi (x)$ is a column with components $\psi _{n}(x)$, one obtains
\begin{equation}
\psi'(x')=L(U^{-1})\psi (x),  \label{tens.9}
\end{equation}
i.e., the transformation law of a spin-tensor field in Minkowski space. This
law corresponds to the representation of the Poincar\'{e} group acting in a
linear space of tensor fields as follows $T(g)\psi (x)=L(U^{-1})\psi (\Lambda^{-1}(x-a))$.
According to (\ref{tens.8}) and (\ref{tens.9}), the functions $\phi (z)$ and
$\psi (x)$ transform according to contragradient representations of the Lorentz
group (we recall that the representation $[T(g^{-1})]^{T}$ is called
contragradient to $T(g)$ \cite{BarRa77}).

For example, let us consider scalar functions on the Poincar\'{e} group
$f_{1}(x,z)=\psi _{\alpha }(x)z^{\alpha }$ and $f_{2}(x,z)=\bar{\psi}_{\alpha }(x)\cc z^{\alpha }$,
which correspond to spinor representations of Lorentz group.
According to (\ref{tens.7}) and (\ref{tens.9})
\begin{equation}
\psi _{\alpha }'(x')=U_{\alpha }^{\;\;\beta }\psi _{\beta}(x),\quad
\bar{\psi}_{\dot{\alpha}}'(x')=\cc U_{\dot{\alpha}}^{\;\;\dot{\beta}}\bar{\psi}_{\dot{\beta}}(x).
\end{equation}
The product $\psi _{\alpha }(x)\bar{\psi}^{\ast \alpha }(x)$ is Poincar\'{e}-invariant.

Thus, \textit{tensor fields of all spins are contained in the decomposition
of the field (\ref{tens.1}) on the Poincar\'{e} group, and the problems of
their classification and construction of explicit realizations are reduced
to the problem of a decomposition of the left GRR}.

The field $f(x,z)$ itself may be regarded as \textit{the generating function}
of usual multi-component spin-tensor fields; the latter arise as the
coefficients of a series in the powers of the orientation (spin) variables $z$.
Below we write out generating function (\ref{fDir}) for spin 1/2
and generating function (\ref{rwe.DK}) for spin 1 (see also \cite{GitSh01} where
some other examples for 2+1 and 3+1 dimensions are contained).

Notice that we have rejected the phase transformations $U=e^{i\phi }$. These
transformations of the $U(1)$ group do not change the space-time coordinates
$x$, but change the phase of $z$. According to (\ref{tens.8}) and (\ref{tens.9}),
this leads to a phase transformation of the tensor field components $\psi_{n}(x)$.
Taking account of this transformations implies a consideration of functions
on the group $T(4)\ctimes \mathrm{Spin}(3,1)\times U(1)$.

\section{The maximal set of commuting operators}

Let us construct the maximal set of commuting operators, which will be used
afterwards to classify fields on the group. Using the above
parameterization, we obtain
\begin{eqnarray}
&&T_{L}(g)f(x,z)=f(g^{-1}x,\;g^{-1}z),\;\;g^{-1}x\leftrightarrow
U^{-1}(X-A)(U^{-1})^{\dagger },\;\;g^{-1}z\leftrightarrow U^{-1}Z,
\label{gen.01} \\
&&T_{R}(g)f(x,z)=f(xg,\;zg),\quad xg\leftrightarrow X+ZAZ^{\dagger },\quad
zg\leftrightarrow ZU.  \label{gen.02}
\end{eqnarray}

In view of (\ref{gen.01}), $x$ transforms according to the vector
representation of the Lorentz group, whereas $z$ transforms according to the
spinor representation. If we restrict the consideration to functions that do
not depend on $z$, then (\ref{gen.01}) reduces to the transformations of the
left quasi-regular representation, that corresponds to the case of a usual
scalar field $f'(x')=f(x)$. If, however, we restrict the
consideration to functions that do not depend on $x$, then (\ref{gen.01})
reduces to the transformations of the left GRR of the Lorentz group.

The generators that correspond to translations and rotations have the form
\begin{eqnarray}  \label{gen.L}
&&\hat{p}_{\mu }=i\partial /\partial x^{\mu }, \quad
 \hat{J}_{\mu\nu }= \hat{L}_{\mu\nu }+ \hat{S}_{\mu \nu},
\\  \label{gen.R}
&&\hat{p}_{\mur }^R=-v^\nu_{\;\;\mur} \hat p_\nu, \quad
\hat{J}_{\mur\nur }^R=\hat{S}_{\mur\nur }^R,
\end{eqnarray}
where $\hat{L}_{\mu \nu }=i(x_{\mu }\partial _{\nu }-x_{\nu }\partial _{\mu})$ are the
operators of projections of the orbital momentum; $\hat{S}_{\mu \nu }$ are the operators
of spin projections, whereas $v_{\;\;\mur}^{\nu }\in SO_{0}(3,1)$ is expressed in terms
of $z$; see (\ref{reg.v0}). The operators of right translations can also be expressed
in the form ${\hat{P}}^{R}=-Z\hat{P}Z^{\dagger }$; the operators $\hat{S}_{\mu \nu }$ and
$\hat{S}_{\mur\nur}^{R}$ are he left and right generators of $SL(2,C)=\mathrm{Spin}(3,1)$
and depend only on $z$ and $\partial /\partial z$,
\begin{eqnarray}   \label{gen.SL}
\hat S_{\mu\nu}&=&i\left(
      (\sigma_{\mu\nu})^{\;\;\beta}_\alpha z^\alpha_{\;\;\alphar} \partial_\beta^{\;\;\alphar} +
      (\bar\sigma_{\mu\nu})^{\dot\alpha}_{\;\;{\dot\beta}} \cc z_{\dot\alpha}^{\;\;\dot{\alphar}} \partial^{\dot\beta}_{\;\;\dot{\alphar}}\right) ,
\\   \label{gen.SR}
\hat S_{\mur\nur}^R &=&i\left(
      (\sigma_{\mur\nur})_{\;\;\betar}^\alphar z^\alpha_{\;\;\alphar} \partial_\alpha^{\;\;\betar} +
      (\bar\sigma_{\mur\nur})_{\dot\alphar}^{\;\;\dot\betar} \cc z_{\dot\alpha}^{\;\;\dot{\alphar}} \partial^{\dot\alpha}_{\;\;\dot{\betar}}\right) ,
\end{eqnarray}
where
\begin{equation}\label{sigmn}
(\sigma_{\mu\nu})_\alpha^{\;\;\beta}
= \frac 14 (\sigma_\mu\bar\sigma_\nu-\sigma_\nu\bar\sigma_\mu)_\alpha^{\;\;\beta},\quad
(\bar\sigma_{\mu\nu})^{\dot\alpha}_{\;\;{\dot\beta}}
= \frac 14 (\bar\sigma_\mu\sigma_\nu-\bar\sigma_\nu\sigma_\mu)^{\dot\alpha}_{\;\;{\dot\beta}}.
\end{equation}
To present the spin operators of the group $M(3,1)$, it is also useful to
apply the three-dimensional vector notation
$\hat S_k =\frac 12\epsilon_{ijk} \hat S^{ij}$, $\hat B_k =\hat S_{k0}$,
\begin{eqnarray}
\hat S_{k}&=& \frac 12 \left( z_{\alphar}\sigma_k \partial^{\alphar}
- \cc z_{\dot{\alphar}} \cc\sigma_k \cc\partial^{\dot{\alphar}}\right), \quad
\\
\hat B_{k}&=& \frac i2 \left( z_{\alphar}\sigma_k \partial^{\alphar}
+ \cc z_{\dot{\alphar}} \cc\sigma_k \cc\partial^{\dot{\alphar}}\right),
\quad
z_{\alphar}=(z^1_{\;\;\alphar} \; z^2_{\;\;\alphar}), \quad
\partial^{\alphar}=(\partial/\partial z^1_{\;\;\alphar} \; \partial/\partial z^2_{\;\;\alphar})^T;
\\
\hat S_{k}^R&=& - \frac 12 \left( z^{\alpha}\cc\sigma_k \partial_{\alpha}
- \cc z^{\dot{\alpha}} \sigma_k \cc\partial_{\dot{\alpha}} \right), \quad
\\
\hat B_{k}^R&=& - \frac i2 \left( z^{\alpha}\cc\sigma_k \partial_{\alpha}
+ \cc z^{\dot{\alpha}} \sigma_k \cc\partial_{\dot{\alpha}}\right),
\quad
z^{\alpha}=(z^\alpha_{\;\;\underline{1}} \; z^\alpha_{\;\; \underline{2}}), \quad
\partial_{\alpha}=(\partial/\partial z^\alpha_{\;\;\underline{1}} \; \partial/\partial z^\alpha_{\;\;\underline{2}})^T.
\end{eqnarray}

The algebra of generators (\ref{gen.L}) has the form
\begin{eqnarray}
&&[\hat{p}_{\mu },\hat{p}_{\nu }]=0,\quad
[\hat{J}_{\mu\nu},\hat{p}_{\rho}]=i(\eta_{\nu \rho}\hat{p}_{\mu} -\eta_{\mu \rho}\hat{p}_{\nu}), \quad
\notag
\\
&&[\hat J_{\mu \nu },\hat J_{\rho \sigma }]
=i\eta _{\nu \rho }\hat J_{\mu\sigma } -i\eta _{\mu \rho }\hat J_{\nu \sigma }
-i\eta _{\nu \sigma}\hat J_{\mu \rho} +i\eta _{\mu \sigma }\hat J_{\nu \rho }\;.  \label{komm0}
\end{eqnarray}
The right generators (\ref{gen.R}) obey the same commutation relations.
Since the group multiplication is associative, $(g_1^{-1}h)g_2=g_1^{-1}(hg_2)$,
each of the right generators (\ref{gen.R}) commutes with each of the left
generators (\ref{gen.L}).

In the space of Fourier transforms
\begin{equation}
\varphi (p,z)=(2\pi )^{-d/2}\int f(x,z)e^{ipx}dx,
\end{equation}
is not difficult to obtain analogies of formulas (\ref{gen.01}), (\ref{gen.02})
in the momenta representation:
\begin{eqnarray}
&&T_{L}(g)\varphi (p,z) = e^{iap^{\prime}} \varphi (p^{\prime},g^{-1}z),
  \quad p^{\prime}=g^{-1}p \leftrightarrow P^{\prime}=U^{-1}P(U^{-1})^\dagger,
  \quad P=p_\mu \sigma^\mu , \quad     \label{Lregp}
\\
&&T_{R}(g)\varphi (p,z) = e^{-ia'p} \varphi (p, z g), \quad
  a'\leftrightarrow A'=ZAZ^{\dagger}.   \label{Rregp}
\end{eqnarray}
Note that $\det Z$ and $\det P=m^{2}$ are invariant with respect to
transformations (\ref{Lregp}), (\ref{Rregp}). Here, $m^{2}$ is an eigenvalue
of the Casimir operator $\hat{\mathrm{p}}^{2}=\eta ^{\mu \nu }p_{\mu }p_{\nu}$.

In the Euclidean case (group $M(3)$), one has two kinds of representations,
depending the value of the squared momentum $\det P=\mathrm{p}^{2}$:
1) $\mathrm{p}^{2}\neq 0$ (moving particles);
2) $\mathrm{p}^{2}=0$ (particles at rest); in this case $p_{i}=0$, and the irreps
are labeled by the eigenvalues of the Casimir operator of the rotation subgroup.

For the group $M(3,1)$, there exist four kinds of transformations, depending
the eigenvalues of $m^{2}$ of the Casimir operator: 1) $m^{2}>0$ (massive
particles); 2) $m^{2}<0$ (tachyons); 3) $m^{2}=0$, $p_{0}\neq 0$ (massless
particles); 4) $m^{2}=p_{0}=0$; the irreps are labeled by the eigenvalues of
the Casimir operator for the Lorentz subgroup; besides, the corresponding
functions are independent of $x$.

To classify functions $f(x,z)$ on the Poincar\'{e} roup, we use the maximal
set of commuting operators. In accordance with the theory of harmonic
analysis on Lie groups \cite{ZhelSc83,BarRa77}, there exists a maximal set
of commuting operators, which includes Casimir operators, a set of the
left generators and a set of the right generators (both sets contain the same
number of generators). The total number of commuting operators is equal to
the number of parameters of the group. In a decomposition of the left GRR
(\ref{gen.01}) the nonequivalent representations are distinguished by
eigenvalues of the Casimir operators, equivalent representations are
distinguished by eigenvalues of the right generators, and the states inside
the irrep are distinguished by eigenvalues of the left generators. And, vice
versa, in an expansion of the right GRR (\ref{gen.02}) the states inside an
irrep are distinguished by the right generators, while equivalent irreps are
distinguished by the left generators.

Notice that some aspects of the theory of harmonic analysis on the 3+1 and
2+1 Poincar\'{e} groups have been considered in
\cite{Ridea66,Hai,Varla04,Varla07} and \cite{GitSh97} respectively.

Functions on the Poincar\'{e} group $M(3,1)$ depend on $10$ parameters, and
correspondingly, there exist $10$ commuting operators (two Casimir operators and
two sets of four operators, constructed from the left (\ref{gen.L}) and right
(\ref{gen.R}) generators).

The Poincar\'{e} group and the (spinorial) Lorentz subgroup have two Casimir
operators each:
\begin{align}
&\hat {\rm p}^2 =\hat p_\mu \hat p^\mu, \quad
 \hat {\rm W}^2 =\hat W_\mu \hat W^\mu, \quad  \hbox{ where} \quad
 \hat W^\mu=\frac 12\epsilon^{\mu\nu\rho\sigma}\hat p_\nu \hat J_{\rho\sigma}
 = \frac{1}{2}\epsilon^{\mu\nu\rho\sigma}\hat p_\nu \hat S_{\rho\sigma},
\label{gen.Pcas}
\\
&\frac 12 \hat S_{\mu\nu}\hat S^{\mu\nu}=
  \frac 12 \hat S^R_{\mu\nu}\hat S_R^{\mu\nu}=
  \hat {\bf S}^2 -\hat {\bf B}^2, \quad
 \frac 1{16}\epsilon^{\mu\nu\rho\sigma}\hat S_{\mu\nu}\hat S_{\rho\sigma}
 =\frac 1{16}\epsilon^{\mu\nu\rho\sigma}\hat S^R_{\mu\nu}\hat S^R_{\rho\sigma}
 =\hat{\bf S}\hat{\bf B}.
\label{gen.Lcas}
\end{align}
Here, we shall examine a set of $10$ commuting operators on the group $M(3,1)$:
\begin{equation}
\hat p_\mu,\; \hat {\rm W}^2,\; \hat{{\bf p}}\hat{{\bf S}}\;
(\hat S_3\hbox{ in the rest frame}), \;
\hat {{\bf S}}^2 -\hat {{\bf B}}^2, \; \hat{{\bf S}}\hat{{\bf B}}, \; \;
\hat S_3^R,\; \hat B_3^R, \; \;
\label{gen.31set}
\end{equation}
including four left generators $\hat{p}_{\mu }$ (the eigenvalue of the
Casimir operator $\hat{\mathrm{p}}^{2}$ obviously is expressed in terms of
their eigenvalues), the Lubanski--Pauli operator $\hat{W}^{2}$, and the
operator of helicity $\hat{\mathbf{p}}\hat{\mathbf{S}}$, expressed in terms
of the left generators. Besides the Casimir operators and functions of the
left generators, the set also includes four functions of the right generators:
$\hat{\mathbf{S}}^{2}-\hat{\mathbf{B}}^{2},\;
\hat{\mathbf{S}}\hat{\mathbf{B}},\;\hat{S}_{3}^{R},\;\hat{B}_{3}^{R}$.
The first two of them are the Casimir
operators of the Lorentz group and determine the characteristics $j_{1},j_{2}$
of the irreps $T_{[j_{1}j_{2}]}$ of this group. Here,
$\hat{\mathbf{S}}^{2}=\hat{S}^{k}\hat{S}_{k}=\hat{S}_{R}^{k}\hat{S}_{k}^{R}$,
$\hat{\mathbf{B}}^{2}=\hat{B}^{k}\hat{B}_{k}=\hat{B}_{R}^{k}\hat{B}_{k}^{R}$.

It is known that $\hat S_k$ and $\hat B_k$ can be linearly combined into
$\hat M_k$ and $\hat{\bar M}_k$,
\begin{eqnarray}
&&\hat M_k=\frac 12(\hat S_k-i\hat B_k)=\frac 12z^\alphar\sigma_k\partial_\alphar, \quad
\nonumber \\
&&\hat {\bar M}_k=-\frac 12(\hat S_k+i\hat B_k)=\frac 12\cc z^{\dot\alphar} \cc\sigma_k\partial_{\dot\alphar}, \quad
\label{gen.MN}
\end{eqnarray}
so that $[\hat M_i,\hat {\bar M}_k]=0$; in addition, unitary representations
of the Lorentz group, according to the condition $\hat S_k^\dagger =\hat S_k$,
$\hat B_k^\dagger =\hat B_k$, must satisfy the relation $\hat M_k^\dagger = \hat {\bar M}_k$
(for finite-dimensional non-unitary representations $\hat S_k^\dagger =\hat S_k$,
$\hat B_k^\dagger =-\hat B_k$ and $\hat M_k^\dagger =-\hat{\bar M}_k$).

Taking into account the fact that $\hat{M}_{k}$ and $\hat{\bar{M}}_{k}$ obey
the commutation relations of the algebra $su(2)$, we find the spectrum of
the Casimir operators of the Lorentz subgroup as follows:
\begin{eqnarray}
&&\hat {\bf S}^2 -\hat {\bf B}^2 = 2(\hat {\mathbf M}^2+\hat {\bar{\mathbf M}}^2)
  =2j_1(j_1+1)+2j_2(j_2+1)=-\frac 12 (k^2-\rho ^2 -4), \quad
\nonumber \\
&&\hat {\bf S}\hat {\bf B} = -i(\hat {\mathbf M}^2-\hat {\bar{\mathbf M}}^2)
  =-i\left( j_1(j_1+1)-j_2(j_2+1)\right) =k\rho,\qquad
\nonumber \\
&&\hbox{where } \quad \rho=-i(j_1+j_2+1), \quad  k=j_1-j_2.
\label{gen.Lcas1}
\end{eqnarray}
Therefore, irreps of the Lorentz group $SL(2,C)$ are labeled by a pair of
numbers, $[j_{1},j_{2}]$. It is convenient to label unitary
infinite-dimensional irreps by pairs of numbers $(k,\rho )$; besides the
irreps $(k,\rho)$ and $(-k,-\rho)$ are equivalent \cite{BarRa77,GelGrV66}.

In the conventional description of relativistic particles (in $3+1$ dim.) in
terms of spin-tensor fields, based, in particular, on a classification of
Poincar\'{e} and Lorentz group representations, there appear 8 particle
characteristics (quantum numbers): 2 numbers $j_{1}$ and $j_{2}$ that label
Lorentz group representations and 6 numbers related to the Poincar\'{e} group,
those are the mass $m$, spin $s$ (Casimirs of the group), and 4 numbers,
which are eigenvalues of some combinations of left generators of the group.
In particular, the latter 4 numbers can be some components of the momentum
and a spin projection.

In the above proposed description of relativistic oriented objects, based on
a classification of group representations of transformation of both s.r.f.
and b.r.f., an oriented object is characterized already by 10 quantum
numbers. These numbers are related to a set of commuting operators
(\ref{gen.31set}) that are combinations of the left and right generators of the
Poincar\'{e} group. Here, in addition to the 2 Casimirs and 4 left generators of
the group, we have 4 numbers which are eigenvalues of some combinations of
right generators of the Poincar\'{e} group. Among the latter, we have 2 Casimirs
of the (right) Lorentz subgroup that fix $j_{1}$ and $j_{2}$.

Thus, as compared with the conventional description, in the proposed one, an
oriented object has two more characteristics. Their physical interpretation
is a subject of an individual investigation.

In addition to the set of commuting operators (\ref{gen.31set}), we use the
generator of phase transformations of $Z$ (\ref{u1_q}), which are also
symmetry transformations of field on the Poincar\'{e} group,
\begin{equation}
\hat{\Phi}=-i\partial /\partial \phi =l\hat{\Gamma}^{5},
\end{equation}
where $l$ labels the irrep of $U(1)$ that governs the transformation of $Z$,
the chirality operator being
\begin{equation} \label{gen.gamma5}
\hat \Gamma^5 = {\textstyle \frac{1}{2}} \left( z^{\alpha}_{\;\;\alphar} \partial_{\alpha}^{\;\;\alphar} -
 \cc z^{{\dot\beta}}_{\;\;\dot\betar} \partial_{{\dot\beta}}^{\;\;\dot\betar} \right).
\end{equation}
In the irreps of the Lorentz group $T_{[j_{1}j_{2}]}$ its eigenvalue is
$\Gamma ^{5}=j_{1}-j_{2}$. It anticommutes with the operator of
space reflection $P$, and, therefore, its eigenvalues change their sign
under the action of $P$.

\section{Fields on the Poincar\'e group. Symmetries}

In the general case, symmetry transformations of a certain object are given by
some transformations that leave this object invariant. For physical systems,
such transformations are usually regarded as symmetries of a Lagrangian (or
those of a Hamiltonian; in this case one separately examines dynamical
symmetries, including transitions between the states with different values of
energy) of this system and of the corresponding equations.

In a broad sense, a symmetry operator of an equation is an arbitrary operator
$\hat{B}$ that transforms solutions $\psi$ of an equation $\hat{L}\psi=0$ into
solutions $\hat{B}\psi$ of the same equation, i.e, an operator that obeys the
condition
\begin{equation}
\hat{L}(\hat{B}\psi )=0  \label{symm.01}
\end{equation}
for each $\psi$ that belong to the totality of solutions of this equation
(see, for example, \cite{FusNi94}).

In what follows, it is expedient to specify the class of symmetry operators
under consideration (e.g., linear, non-linear, integral, integro-differential,
and so on). Of special interest are symmetry operators that belong to the
class of linear differential operators of first order, and which may be
considered as generators of continuous groups of transformations. These are
so-called Lie, or classical, symmetries; ``higher'' (non-classical) and
non-local symmetries are a subject of a separate study (see \cite{FusNi94} and
references therein).

Free relativistic wave equations are invariant with respect to the group of
relativistic symmetry, the Poincar\'{e} group. In essence, they are conditions
that the corresponding field $\psi$ should belong to a certain representation
of this group. For instance, the Dirac equation can be obtained as a condition
of belonging to an irrep of the extended (due to discrete transformations)
Poincar\'{e} group, which is characterized, in particular, by the sign of the
product of parity, charge, and energy (an exact formulation is given in
\cite{BucGiS02}). Symmetry transformations of an equation, evidently, map a
field carrying some representation into a field carrying the same
representation. This observation indicates the possibility to define the
notion of field symmetry without appealing to any relativistic wave equations.

So, let a field $\psi$ be transformed according to the representation $T(g)$
of the group $G$. Let us call the operator $\hat{\mathbf{B}}$
\textit{a symmetry operator of this field} (in a broad sense), in case the field
$\hat{\mathbf{B}}\psi$ is transformed according to the same representation.

Consider symmetry transformations using the example of a scalar field on a
certain homogenous space of the group $G$. The representation in the space
of scalar functions on the homogenous space $G/K$, $K\in G$ (the so-called
left quasi-regular representation) is given by
\begin{equation}
T(g)f(y)=f'(y)=f(g^{-1}y),\quad g\in G,\quad y\in G/K,  \label{QRRL}
\end{equation}
or $f'(y')=f(y)$, where $y'=gy$. Let us denote
the space of scalar functions in $G/K$ as $V_{G/K}$.

If the subgroup $K$ is trivial, we have an important particular case. In this
case, the coset space is identical with the group $G$ itself and we deal with
the GRR (\ref{reg.4}), and, correspondingly, with a scalar field on the entire
group (\ref{reg.5}).

According to the above definition, \textit{symmetry transformations of the
scalar field on} $G/K$ are mappings of a field into itself:
\begin{equation}
f(y)\in V_{G/K}\rightarrow \hat{\mathbf{B}}[f(y)]\in V_{G/K}.
\label{symm.02}
\end{equation}
We shall impose the following natural restrictions on $\hat{\mathbf{B}}$:

\noindent 1) the operator $\hat{\mathbf{B}}$ is invertible;

\noindent 2) the transformation $\hat{\mathbf{B}}[\psi(y)]$ reduces to a change
of function arguments on a homogenous space (change of coordinates)
$\hat{\mathbf{B}}(f(y))=f(\tilde{y})$.

That is, $\tilde y=\hat By$, where $\hat B$ acts on $G/K$, and, due to the
existence of $\hat B^{-1}$, the mapping $y\leftrightarrow \tilde y$ is
bijective.

Acting by the operator $\hat{\mathbf{B}}$ on (\ref{QRRL}), we obtain
\begin{eqnarray}
&&\hat{\mathbf{B}}T(g)\hat{\mathbf{B}}^{-1}\hat{\mathbf{B}}\psi (y)
=\hat{\mathbf{B}}\psi (g^{-1}y),
\label{symm.03}
\\
&&\hat{\mathbf{B}}T(g)\hat{\mathbf{B}}^{-1}\psi (\tilde{y})
=\psi (\hat{B}g^{-1}\hat{B}^{-1}\tilde{y}),\quad
\hat{\mathbf{B}}T(g)\hat{\mathbf{B}}^{-1}=T(\tilde{g}),
\notag
\end{eqnarray}
where $T(\tilde{g})f(\tilde{y})=f(\tilde{g}^{-1}\tilde{y})$, and the mapping
$g\rightarrow\tilde{g}=\hat{B}g\hat{B}^{-1}$ preserves the law of group
composition, and, therefore, determines an automorphism of the group $G$.

One can say that the symmetry transformations of the field $f(y)\to f(\tilde y)$
are generated by these automorphisms.

Let us consider the Poincar\'{e} group $M(3,1)$. From the physical point of
view, it is interesting to consider the special case of spaces that include
the Minkowski space as their subspace. Notice that there exist $11$ spaces of
this kind \cite{Finke55}.

First of all, let us examine a scalar field $f(x)$ in the Minkowski space,
$x\in M(3,1)/\mathrm{Spin}(3,1)$. The inner automorphisms
\begin{equation}
g\rightarrow g_{0}gg_{0}^{-1},\quad x\rightarrow g_{0}x
\label{symm.04}
\end{equation}
correspond to the left finite transformations of the Poincar\'{e} group
(proper Lorentz transformations), i.e., to changes of the s.r.f..
The outer automorphisms of the Poincar\'{e} group
\begin{equation}
g\rightarrow \hat{B}g\hat{B}^{-1},\quad x\rightarrow \hat{B}x
\label{symm.05}
\end{equation}
correspond to space- and time-reflections, as well as to scale transformations (dilatations).

A scalar field $f(q)$ on the Poincar\'{e} group, $q\in M(3,1)$, is a special
case. In comparison with the case of homogenous spaces $G/K$ with a nontrivial
subgroup $K$, in this case there exist a larger variety of symmetries. Namely,
one can multiply $q$ by an element of the group not only from the left but
also from the right, and, therefore, we consider a representation
$\mathbb{T}(g,h)$ of the direct product $M(3,1)\times M(3,1)$,
$\mathbb{T}(g,h)f(q)=f(g^{-1}qh)$.

Symmetry transformations of the general form correspond to the automorphisms
\begin{equation*}
(g,h)\rightarrow (\tilde{g}=B_{1}gB_{1}^{-1},\tilde{h}=B_{2}hB_{2}^{-1})
\end{equation*}
of the group $M(3,1)\!\times\! M(3,1)$. They are generated by three kinds of
transformations of $q$:
\begin{eqnarray}
&&g\to g_0gg_0^{-1}, \quad h\to h, \quad  q\to g_0q \quad
\hbox{(proper Lorentz transformations)};
\label{lab_q} \\
&&g\to g, \quad h\to h_0hh_0^{-1}, \quad q\to qh_0^{-1} \quad
\hbox{(right transformations)};
\label{lok_q} \\
&&g\to \hat B g\hat B^{-1}, \quad h\to \hat B h\hat B^{-1}, \quad q\to \hat B q\hat B^{-1} \quad
\hbox{(outer automorphisms)}.
\label{out_q}
\end{eqnarray}
In comparison with the symmetries of a scalar field in the Minkowski space
(\ref{symm.04}), (\ref{symm.05}), we have additional symmetries (\ref{lok_q}).

Instead of the right transformations (\ref{lok_q}) corresponding to a change
of the s.r.f., we can consider inner automorphisms being a composition of left
and right transformations,
\begin{equation}
g\rightarrow h_{0}gh_{0}^{-1},\quad h\rightarrow h_{0}hh_{0}^{-1},\quad
q\rightarrow h_{0}qh_{0}^{-1}\quad \hbox{(inner automorphisms)}.
\label{inn_q}
\end{equation}

Therefore, a scalar field on the group, besides the symmetry with respect to
the proper Lorentz transformations, also possesses nontrivial symmetries
related to inner $q\rightarrow h_{0}qh_{0}^{-1}$ and outer (space and time
reflections $I_{s}$,$I_{t}$, as well as scale transformations
$(A,U)\rightarrow (cA,U)$, $(X,Z)\rightarrow (cX,Z)$) automorphisms. Notice
that, unlike the proper Lorentz transformations, the field symmetries
corresponding to automorphisms, generally do not preserve the interval (in
particular, this is valid for the scale transformations).

The symmetries (\ref{lab_q}) correspond to a change of the s.r.f., whereas
(\ref{lok_q}) correspond to a change of the b.r.f.. Outer automorphisms
(\ref{out_q}), however, act simultaneously on both reference frames. It is
necessary to obtain a group element $\tilde{q}\in G$ as a result of the
transformation. Let us explain this using the example of a rotator, where the
matrix $V\in SO(3)$ relates s.r.f. and b.r.f., see (\ref{rot.1}),(\ref{rot.2}).
As a result of a space reflection of one of the frames, we find that they
are now related by a matrix $V'$, $\det V'=-1$, and, therefore, $V'\notin SO(3)$.

If one explicitly indicates $i$ as an argument of functions on the group,
then complex conjugation
\begin{equation}
f(x,z,i)\rightarrow f(x,\cc z,-i)  \label{con_q}
\end{equation}
can also be considered as a change of arguments of a function. As will be seen below,
this corresponds to charge conjugation (see also \cite{GitSh01,BucGiS02}).

The phase transformations $Z'=e^{il\varphi }Z$ that do not affect
$x^{\mu }$ can also be considered as symmetry transformations of a field;
however, on a larger space, i.e., fields on the group $M(3,1)\times U(1)$
(instead of the unimodularity condition $\det Z=1$, one has $|\det Z|=1$),
\begin{equation}
z\rightarrow ze^{il\varphi }.  \label{u1_q}
\end{equation}
A real-valued number $l$ determines one-dimensional irreps of the group $U(1)$.
These transformations commute with transformations from $M(3,1)$, so that
the order of multiplication is not important, and, therefore, these
transformations can be treated as both (\ref{lab_q}) and (\ref{lok_q}).
However, in view of the fact that, as distinct from the Lorentz
transformations (\ref{lab_q}), they do not affect $x^{\mu}$, it is natural to
consider them as a particular case of the right transformations (\ref{lok_q}),
where $g_{0}\in M(3,1)\times U(1)$.

A scalar field on the Poincar\'{e} group $f(x,z)$ has a larger variety of
symmetries. Indeed, let us examine the action of right transformations on a
usual scalar field $f(x)$. Right rotations and boosts, in accordance with
(\ref{reg.13}), do not affect $x$, i.e., they are reduced to the identity
transformation. Right translations are not symmetry transformations, since,
in accordance with (\ref{reg.13}), they map functions $f(x)$ on a homogenous
space into functions $f(x,z)$ on the entire group. Further, in the case of
fields carrying a fixed nonzero mass, symmetries do not include scale
transformations which are a particular case of automorphisms (\ref{out_q}),
and so on.

The physical sense of the right transformations (\ref{lok_q}) can be clarified
by the example of the compact subgroup of rotations $SO(3)\sim SU(2)$, which
describes a rotator. According to the above consideration, in the case of a
nonrelativistic rigid rotator left transformations (changes of s.r.f.)
correspond to external symmetries (symmetries of the embedding space), whereas
right transformations (changes of b.r.f.) correspond to symmetries of the
rotating body itself. However, if the body is not symmetric, then only one
part of the right symmetries of a field on the group $SU(2)$ corresponds to
the symmetries of the rotator Hamiltonian (\ref{rot.11}), while the remaining
part of the right symmetries are violated. In a similar way, the
transformations (\ref{lab_q}) of a scalar field on the Poincar\'{e} group,
i.e., the proper Lorentz transformations, correspond to external symmetries,
and one can suppose that (\ref{lok_q}) corresponds to internal symmetries.
Below, we present a more detailed analysis.

\section{Discrete symmetries}

A particular case of symmetries of a scalar field on the Poincar\'{e} group
is given by discrete symmetry transformations, which cannot be related
continuously with the identity transformation. These transformations are
generated by two outer automorphisms of the type (\ref{out_q}): the space
$I_{s}$ (also denoted by $P$, as the transformation of space parity) and the
time $I_{t}$ reflections, as well as the complex conjugation of functions on
the group (\ref{con_q}), which corresponds to charge conjugation $C$.

In fact, one has to consider a larger class of transformations, namely
compositions of the above-mentioned outer automorphisms and internal, or
equivalently, some right transformations (namely, right rotations) of the
Poincar\'{e} group. It is precisely these combinations that present the
symmetry transformations of various important Hamiltonians and relativistic
wave equations. In other words, outer automorphisms and right transformations
must be examined simultaneously.

Therefore, it is expedient to approach the definition of discrete
symmetries from a different point of view. Let us define them as such
symmetry transformations of a scalar field on the Poincar\'e group that
satisfy the following conditions:

(i) A squared discrete transformation equals to the identity transformation;

(ii) The totality of discrete transformations forms a group.

From (ii), it follows that a product of discrete transformations is also a
discrete transformation and its square equals to the identity operator.
Hence follows that discrete transformations commute with each other.

Among the proper Lorentz transformations (\ref{lab_q}), condition (i) is
satisfied only by the rotation by the angle $2\pi$. Condition (i) is also
met by transformations generated by involutive (both outer (\ref{out_q}) and
inner (\ref{inn_q})) automorphisms of the group. However, it is only the
subset of the indicated automorphisms that satisfy condition (ii). Finally,
the above definition is also met by the complex conjugation (\ref{con_q}) of
functions on the group.

In \cite{BucGiS02}, it was shown that a field on $M(3,1)$ has only six
independent transformations that obey the above definition, namely, the
rotation by $2\pi$, the complex conjugation of functions on the group, two
outer automorphisms (the space reflection $I_s$ and the time reflection $I_t$),
and two inner automorphisms.

Automorphisms determine transformations of the space-time and orientation
coordinates $(x,z)$. A substitution of the transformed coordinates into
functions $f(x,z)$ (or generators (\ref{gen.L}), (\ref{gen.R})) leads to a
change of the sign of some physical quantities. (Notice that a simultaneous
change of variables in the expressions for the generators and functions
$f(x,z)$ leaves the signs unchanged.)

Let us start from outer automorphisms. In the case of space reflection $I_{s}$:
\[
e_{\mu}\rightarrow-(-1)^{\delta_{0\mu}}e_{\mu},\quad
\xi_\mur\rightarrow-(-1)^{\delta_{0\mur}}\xi_\mur ,
\]
or in terms of the $2\times2$ matrices:
\[
E=\sigma^{\mu}e_{\mu}\rightarrow\sigma^{2}E\sigma^{2}, \quad
\Xi=\sigma^\mur \xi_\mur \rightarrow \sigma^{2}\Xi\sigma^{2}.
\]
Hence, for coordinates of a vector $x=x^{\mu}e_{\mu}$ we have
$x^{\mu}\rightarrow-(-1)^{\delta_{0\mu}}x^{\mu}$, or $X\rightarrow\overline{X}$,
whereas the orientation variables, according to (\ref{par.8}) and (\ref{reg.0}),
are found in the form $Z\rightarrow\sigma^{2}Z\sigma ^{2}=(Z^{\dagger})^{-1}$.
Thus, the space reflection corresponds to an outer automorphism of the Poincar\'{e}
group:
\begin{equation}
I_{s}:\qquad (X,Z)\rightarrow (\overline{X},(Z^{\dagger })^{-1}).  \label{P}
\end{equation}
In the case of space reflection, $x$ and $z$ have to be replaced in all of the
constructions, according to (\ref{P}). In particular, the momentum is found to
be $P\rightarrow\overline{P}$, where $\overline{P}=p_{\mu}\bar{\sigma}^{\mu}$.
The generators of space rotations remain unchanged, whereas the generators of
boosts do change their sign.

In a similar way, in the case of time reflection,
$e_{\mu }\rightarrow (-1)^{\delta _{0\mu }}e_{\mu }$,
$\xi _\mur\rightarrow (-1)^{\delta _{0\mur}}\xi _\mur$, we find
\begin{equation}
I_{t}:\qquad (X,Z)\rightarrow (-\overline{X},(Z^{\dagger })^{-1}).
\label{It}
\end{equation}
The inversion $I_{x}=I_{s}I_{t}$ corresponds to an automorphism
\begin{equation}
I_{x}:\qquad (X,Z)\rightarrow (-X,Z).  \label{Ix}
\end{equation}

One can show that, in the framework of characteristics determined by the
Poincar\'{e} group, complex conjugation (the change $i\rightarrow-i$)
\begin{equation}
C:\qquad f(x,z){\rightarrow}\cc f(x,z), \label{C}
\end{equation}
corresponds to the charge conjugation. Indeed, both (\ref{C}) and the charge
conjugation change signs of all the generators,
$\hat{p}_{\mu}\rightarrow -\hat{p}_{\mu}$,
$\hat{L}_{\mu\nu}\rightarrow-\hat{L}_{\mu\nu}$,
$\hat{S}_{\mu\nu}\rightarrow-\hat{S}_{\mu\nu}$. The study of relativistic wave
equations shows \cite{GitSh01} that transformation (\ref{C}) also changes the
sign of the current $j^{\mu}$.

Time reversal $T$, as well as time reflection $I_{t}$, is determined by the
relation $X\rightarrow-\overline{X}$, which, however, is valid under the
subsidiary condition of the preservation of the sign of energy, which
corresponds to $P\rightarrow\overline{P}$. In consequence, there hold the relations
\[
\hat{p}_{\mu }\rightarrow -(-1)^{\delta _{0\mu }}\hat{p}_{\mu },\quad
\hat{L}_{\mu \nu }\rightarrow -(-1)^{\delta _{0\mu }+\delta _{0\nu }}\hat{L}_{\mu \nu },\quad
\hat{S}_{\mu \nu }\rightarrow -(-1)^{\delta _{0\mu }+\delta _{0\nu }}\hat{S}_{\mu \nu }.
\]
These conditions are met by the transformation $CI_{t}$.

However, it is known \cite{UmeKaT54,KemPoP59} that it is possible to give
two distinct definitions of time-reversal transformation obeying the
above-mentioned conditions. Wigner's time reversal $T_{w}$ leaves the total
charge (and, correspondingly, $j^{0}$) unaltered, and reverses the direction
of the current $j^{k}$. Schwinger's time reversal $T_{sch}$ leaves the current
$j^{k}$ invariant and reverses the charge \cite{Schwi51}. The transformation
$CI_{t}$ changes the sign of $j^{0}$, and, therefore, can be identified with
Schwinger's time reversal, $T_{sch}=CI_{t}$.

Wigner's time reversal $T$ and the $CPT$-transformation can be defined by
considering both outer and inner automorphisms of the proper Poincar\'{e} group
\cite{BucGiS02}. Namely, $CPT=I_{x}I_{z}$, where $I_{z}$ is defined as follows:
\begin{equation}
I_{z}:\qquad (X,Z){\rightarrow }(X,Z(-i\sigma _{2})),  \label{Iz}
\end{equation}
and is a composition of the inner automorphism
$(X,Z)\rightarrow (\overline{X}^{T},(Z^{T})^{-1})$ (which, in its turn, can be
presented as a product of automorphisms (\ref{P}) and $(X,Z)\rightarrow (\cc X,\cc Z)$)
and of a rotation by the angle $\pi $. Wigner's time reversal is a composition
of the above-considered transformations:
\begin{equation}
T=I_{z}T_{sch}=CI_{z}I_{t}.  \label{T}
\end{equation}
(Discrete transformations $I_{z}$ and $T$ are compositions of a transformation
with a unity square and of a rotation by $\pi$ that changes the signs of two
spatial axes. Consequently, instead of (i), we use, in this case, a weaker condition,
namely, that the square of a discrete transformation be equal to the identity
transformation or the rotation by $2\pi$. This definition does not change anything
conceptually; however, it is more convenient technically.)

One can see that $C^{2}=P^{2}=I_{t}^{2}=1$. The operators $I_{z}$ and $I_{3}$,
the latter acts as $(X,Z)\overset{I_{3}}{\rightarrow}(X,Z(-i\sigma _{3}))$, are
products of the involutive inner automorphism and the rotation by $\pi$.
Correspondingly, $I_{z}^{2}=I_{3}^{2}=T^{2}=R_{2\pi}$, where $R_{2\pi}$ is the
operator of a rotation by $2\pi$, which changes the sign of $z$.

Therefore, charge conjugation corresponds to complex conjugation of scalar
functions on the group, whereas the remaining five independent transformations
correspond to a change of the arguments of scalar functions on the group:
\begin{equation}
\begin{array}{|l|rr|rrrr|}
\hline
        & x^0 & {\boldsymbol x} & z^\alpha   & \cc z_{\dot\alpha} & {\underline z}^\alpha &\cc{\underline z}_{\dot\alpha} \\
\hline
R_{2\pi}& x^0 & {\boldsymbol x} &-z^\alpha   &-\cc z_{\dot\alpha} &-{\underline z}^\alpha &-\cc{\underline z}_{\dot\alpha} \\
P=I_s   & x^0 &-{\boldsymbol x} &-\cc{\underline z}_{\dot\alpha} & {\underline z}^\alpha & \cc z_{\dot\alpha} &-z^\alpha \\
I_x     &-x^0 &-{\boldsymbol x} & z^\alpha   & \cc z_{\dot\alpha} & {\underline z}^\alpha &\cc{\underline z}_{\dot\alpha} \\
I_z     & x^0 & {\boldsymbol x} & {\underline z}^\alpha & \cc{\underline z}_{\dot\alpha} & -z^\alpha &-\cc z_{\dot\alpha} \\
I_3     & x^0 & {\boldsymbol x} &-iz^\alpha  & i\cc z_{\dot\alpha} &i{\underline z}^\alpha &-i\cc{\underline z}_{\dot\alpha} \\
\hline
\end{array}
\label{xz}
\end{equation}
For the sake of clarity, we have used the notation that we applied in \cite{GitSh01},
$z^{\alpha }=z_{\;\;\underline{1}}^{\alpha }$, $\cc z_{\dot{\alpha}}=\cc z_{\dot{\alpha}}^{\;\;\dot{\underline{2}}}$,
${\underline{z}}^{\alpha }=z_{\;\;\underline{2}}^{\alpha }$,
$\cc {\underline z}_{\dot{\alpha}}=\cc z_{\dot{\alpha}}^{\;\;\dot{\underline{1}}}$
(the dot over an index duplicates the sign of the complex conjugation of $z$).

The formulas for the transformations $P,C,T$ of four kinds of spinors, denoted usually (see,
\cite{LanLi4}) by $\xi^{c\alpha}$, $\xi^{\alpha}$, $\eta_{\dot{\alpha}}$, $\eta_{\dot{\alpha}}^{c}$,
are identical (on condition that $P^{2}=1$) with those in the case of
$z^{\alpha}$, ${\underline{z}}^{\alpha}$, $\cc z_{\dot {\alpha}}$, $\cc {\underline z}_{\dot{\alpha}}$.

Expanding $f(x,z)$ in the powers of $z$, one can obtain the transformation
laws for spin-tensors of an arbitrary rank, without any reference to the
Dirac equation or some other RWE \cite{BucGiS02}.

In the general case, it is only a part of discrete transformations that
presents a symmetry transformation of RWE, since some RWE fix certain
characteristics that label representations of the extended (by discrete transformations)
Poincar\'e group. In particular, discrete
symmetries of the Dirac and Weyl equations are generated by two unmatched
sets of three operators, respectively, $P,C,T$ and $PC,I_x,T$.

\section{Classification of scalar functions\newline
and equivalent representations}

Among functions on the group, there are functions that transform identically under
the action of a left GRR (for instance, in the case of $M(3,1)$ these are the functions
$f(x^{\mu},z_{\;\;\underline{1}}^{\alpha})$ and
$f(x^{\mu},z_{\;\;\underline{2}}^{\alpha})$). A natural extension of
functions that transform ``identically'' is given by functions that transform by
equivalent representations.

Let us recall that representations $T_{1}(g)$ and $T_{2}(g)$ acting in linear
spaces $L_{1}$ and $L_{2}$, respectively, are equivalent in case there exists
an nonsingular linear operator $A:L_{1}\rightarrow L_{2}$ such that
\begin{equation}
\hat{A}T_{1}(g)=T_{2}(g)\hat{A}.
\end{equation}
If $L_{1}$ and $L_{2}$ are subspaces of the space of functions in $M(3,1)$,
then $\hat{A}T_{1}(g)f_{1}(x,z)=T_{2}(g)\hat{A}f_{1}(x,z)$, and,
respectively,
\begin{equation*}
f_{2}(x,z)=(\hat{A}f_{1})(x,z),
\end{equation*}
where $f_{1}(x,z)\in L_{1}$ and $f_{2}(x,z)\in L_{2}$.

In particular, the left and the right GRR of a Lie group G are equivalent.
The operator
\begin{equation}
(\hat{A}f)(q)=f(q^{-1})  \label{equiv}
\end{equation}
realizes the equivalence \cite{ZhelSc83,Vilen68t}. This is quite natural,
since the left and right transformations are two representations of the same
abstract Lie group. However, in the case under consideration the left and
right transformations have a different geometrical and physical sense, as
transformations of s.r.f. and b.r.f., which are transformed one into another
by the operator $\hat{A}$. Thus, the former retain the distance
(\ref{orient.2}) or the interval (\ref{orient.3}) unchanged, as distinct from
the latter.

Transformation (\ref{equiv}) implies that, starting from the space of functions
$f(q)$, $q\leftrightarrow (X,Z)$ that depend on the coordinates of b.r.f. in
s.r.f., one goes over to \textit{another} space, namely, that of functions
$f(q^{-1})$, $q^{-1}\leftrightarrow(-Z^{-1}X(Z^{-1})^{\dagger },Z^{-1})$
depending on the coordinates of a s.r.f. in b.r.f..

Let us now consider functions from two different subspaces in the space of
functions on the group, which transform identically under the action $T_{L}(g)$
(e.g., the above-mentioned functions
$f(x^{\mu},z_{\;\;\underline {1}}^{\alpha})$ and $f(x^{\mu},z_{\;\;\underline{2}}^{\alpha})$ ).
A question arises if they describe equal states, and, if they do describe different
states, which are physical characteristics that distinguish these states.

Functions on the Poincar\'{e} group depend on the variables
$z_{\;\,\betar}^{\alpha }$ having two kinds of indices: one of them is
related to the transformations of a s.r.f., the other is related to the
transformations of b.r.f.. Fields corresponding to equivalent
subrepresentations of the left GRR \textit{transform equally under the proper
Lorentz transformations} (changes of s.r.f.) and \textit{differently under a
change of b.r.f. or under automorphisms}, in particular, under discrete
transformations (space reflection, time reversal, charge conjugation). They
can be classified with the help of the right generators of the group, which
are contained in the maximal set of commuting operators. Thus, different
equivalent subrepresentations of the left GRR describe, in the general sense,
different physical situations.

Below, we shall use, whenever necessary, the notation $SL(2,C)_{left}$
(Lorentz transformations) and $SL(2,C)_{right}$ for the action from the left
and from the right.

\section{Four kinds of spinors}

It is known that the nonrelativistic spin is described by the group $SU(2)$
and there exists only one kind of spinors; in the relativistic theory there
exist two kinds of spinors (left and right spinors, transforming differently
by boosts and distinguished usually with the help of dotted and undotted
spinor indices). A manifest construction of the extended Poincar\'{e} group
shows that in the relativistic theory with discrete transformations there
exist \textit{four kinds of spinors}, transforming differently by discrete
transformations; see (\ref{xz}). Namely, besides the dotted and un-dotted
spinors that transform into each other by the space reflection $P$, it is
necessary to distinguish the spinors $z_{\;\;\underline{1}}^{\alpha}$ and
$z_{\;\;\underline{2}}^{\alpha}$, that transform into each other by the
$CPT$-transformation.

The same four kinds of spinors $z_{\;\;\underline{1}}^{\alpha}$,
$z_{\;\;\underline{2}}^{\alpha}$, $\cc z_{\dot{\alpha}}^{\;\;\dot {\underline{2}}}$,
$\cc z_{\dot{\alpha}}^{\;\;\dot{\underline{1}}}$, or, using the notation
from \cite{GitSh01}, $z^{\alpha}$, ${\underline{z}}^{\alpha}$,
$\cc z_{\dot{\alpha}}$, $\cc {\underline z}_{\dot{\alpha}}$, appear, in the
course of a simultaneous consideration of left and right finite
transformations of the Lorentz and Poincar\'{e} groups, as columns and rows of
the matrices $Z,\cc Z\in SL(2,C)$,

\begin{eqnarray}
&&
Z=\left( \begin{array}{cc}
 z^1_{\;\;\underline{1}} & z^1_{\;\;\underline{2}} \\
 z^2_{\;\;\underline{1}} & z^2_{\;\;\underline{2}} \\
\end{array} \right) =
\left( \begin{array}{cc}
 z^1 & {\underline z}^1 \\
 z^2 & {\underline z}^2 \\
\end{array} \right), \quad
\\
&&
\cc Z=\left( \begin{array}{cc}
 \cc z^{\dot 1}_{\;\;\dot{\underline{1}}} & \cc z^{\dot 1}_{\;\;\dot{\underline{2}}} \\
 \cc z^{\dot 2}_{\;\;\dot{\underline{1}}} & \cc z^{\dot 2}_{\;\;\dot{\underline{2}}} \\
\end{array} \right) =
\left( \begin{array}{rr}
 -\cc z_{\dot 2}^{\;\;\dot{\underline{2}}} & \cc z_{\dot 2}^{\;\;\dot{\underline{1}}} \\
 \cc z_{\dot 1}^{\;\;\dot{\underline{2}}} & -\cc z_{\dot 1}^{\;\;\dot{\underline{1}}} \\
\end{array} \right) =
\left( \begin{array}{rr}
 -\cc z_{\dot 2} & \cc {\underline z}_{\dot 2} \\
 \cc z_{\dot 1} & -\cc {\underline z}_{\dot 1} \\
\end{array} \right).
\end{eqnarray}
Note that $z_{\;\;\betar}^{\alpha }z_{\alpha }^{\;\;\betar}/2=\det Z=1$.

A pair of spinors ($z_{\;\;\betar}^{\alpha }$, $\cc z_{\dot{\alpha}}^{\;\;\dot\betar}$
or differential operators
$\partial _{\alpha}^{\;\;\betar}=\partial /\partial z_{\;\;\betar}^{\alpha }$,
$\partial _{\;\;\dot\betar}^{\dot{\alpha}}=\partial /\partial \cc z_{\dot{\alpha}}^{\;\;\dot\betar}$)
allows one to construct not only quantities with vector indices of the same kind (e.g., the left
$\hat{S}_{\mu \nu }$ (\ref{gen.SL}) and right $\hat{S}_{\mur\nur}^{R}$
(\ref{gen.SR}) generators), but also quantities with indices of two different types, one being right,
the other left.

First of all, these are the tetrads $v_{\;\;\nur}^{\mu }$ (\ref{reg.v0}), the operators
\begin{equation}
\hat V_{12}\mathstrut^{\mu \nur} = \frac{1}{2}
\bar \sigma ^{\mu{\dot\alpha}\beta} \sigma^{\nur}_{ \;\; \betar\dot\alphar}
\cc z_{{\dot\alpha}}^{\;\;\dot\alphar}\partial_{\beta}^{\;\;\betar},
\qquad
\hat V_{21}\mathstrut^{\mu \nur} = \frac{1}{2}
\sigma^{\mu}_{\;\; \beta{\dot\alpha}} \bar \sigma ^{\nur\dot\alphar\betar}
z^{\beta}_{\;\;\betar}\partial^{\dot\alpha}_{\;\;\dot\alphar},
\label{spi.v12}
\end{equation}
related by the space reflection, and the operator
\begin{equation}
\hat V_{22}\mathstrut^{\mu\nur} = \frac{1}{2}
\sigma^{\mu}_{\;\; \beta{\dot\alpha}} \bar \sigma ^{\nur\dot\alphar\betar}
\partial^{\dot\alpha}_{\;\;\dot{\alphar}}\partial_{\beta}^{\;\;\betar}.
\label{spi.v22}
\end{equation}
Along with the tetrads, these operators can be used to construct
relativistic wave equations. They connect irreps $T_{[j_{1}j_{2}]}$ of the
Lorentz group with different $j_{1},j_{2}$; the operators
$\hat{V}_{12}\mathstrut ^{\mu \nur}$ and $\hat{V}_{21}\mathstrut ^{\mu \nur}$
preserve $j_{1}+j_{2}$, while the operators
$\hat{V}_{11}\mathstrut ^{\mu \nur}=v^{\mu \nur}$ and
$\hat{V}_{22}\mathstrut ^{\mu \nur}$ preserve $j_{1}-j_{2}$.

It should also be noted that the subspaces of functions
$f(x^{\mu },z_{\;\;\underline{1}}^{\alpha },\cc z_{\;\;\dot{\underline{2}}}^{\dot{\alpha}})$,
$f(x^{\mu },z_{\;\;\underline{2}}^{\alpha },\cc z_{\;\;\dot{\underline{1}}}^{\dot{\alpha}})$
are invariant for the operators $\hat{V}_{12}\mathstrut ^{\mu \nur}$ and
$\hat{V}_{21}\mathstrut ^{\mu \nur}$.
Let us denote these subspaces by $V_{-}$ and $V_{+}$. The polynomials of
degree $2s$ in the subspaces $V_{-}$ and $V_{+}$ are eigenfunctions of the
operator of right rotations $\hat{S}_{R}^{3}$ with the eigenvalues $s$ and $-s$
(since $z_{\;\;\underline{1}}^{\alpha },\cc z_{\;\;\dot{\underline{2}}}^{\dot{\alpha}}$
and $z_{\;\;\underline{2}}^{\alpha },\cc z_{\;\;\dot{\underline{1}}}^{\dot{\alpha}}$
are eigenvectors for $\hat{S}_{R}^{3}$ with the respective eigenvalues 1/2 and -1/2).

\section{Left-invariant relativistic wave \newline
equations}

Consider, at first, equations that determine the eigenvalues of the Casimir
operators $\hat{p}^{2}=\hat{p}_{R}^{2}$ and $\hat{W}^{2}=\hat{W}_{R}^{2}$ for
the Poincar\'{e} group, those are the Klein-Gordon equation and the
Lubanski-Pauli equation:
\begin{align}
\hat{p}^{2}f(x,z) &  =m^{2}f(x,z),\qquad  \label{casimir-p2}\\
\hat{W}^{2}f(x,z) &  =-m^{2}s(s+1)f(x,z). \label{casimir-w2}
\end{align}
They must be satisfied by any free fields with a definite mass $m$ and spin $s$.
These equations are invariant with respect to the left (\ref{lab_q}) and
right (\ref{lok_q}) transformations, and also with respect to some of the
outer automorphisms (\ref{out_q}) (namely, the involutive automorphisms)
and the phase transformations (\ref{u1_q}). Amongst all the symmetries of a field
on the Poincar\'{e} group, they do not possess only the symmetry with respect
to the part of the outer automorphisms (\ref{out_q}) (namely, scale transformations).

As known, free relativistic particles are usually described on the basis of
such RWE as Dirac, Weyl, Duffin-Kemmer, etc.
Making a comparison of the latter equations with equations (\ref{casimir-p2}),
(\ref{casimir-w2}), we note two aspects closely related to each other. In the
first place, some equations of first order contain additional information
in comparison with (\ref{casimir-p2}) and (\ref{casimir-w2})
(Lorentz characteristics $j_{1},j_{2}$, chirality or inner parity, charge). For
example, the Dirac equation entails the existence of a pair of particles
related through charge conjugation. Secondly, these equations possess only
some of the symmetries (in particular, discrete) of equations
(\ref{casimir-p2}), (\ref{casimir-w2}).

The point is that the conventional approach to relativistic wave equations
takes into account only the characteristics determined by the left generators
of the Poincar\'e group (these can be used to construct only six commuting
operators: the 4-momentum $\hat p_\mu$, the operator $\hat W^2$, that
determines spin $s$, and the spin-projection $\hat S^3$).

If we, however, take into account not only the left but also the right
transformations, then, instead of two (both dotted and undotted), we now have
four kinds of spinors (two pairs, related by charge conjugation), Lorentz
characteristics being $j_{1},j_{2}$; altogether, there are four additional
quantum numbers (in total, the maximal set of commuting operators contains ten
operators, whose number is equal to that of the parameters of the Poincar\'{e} group).

Correspondingly, equations containing additional information possess
non-trivial transformation properties with respect to the right
transformations and can be consistently deduced from group-theory conditions
only by taking into account the ``right'' characteristics, or, at any rate,
the characteristics of the Poincar\'e group being extended by discrete transformations.

It should be noted that the situation with RWE is analogous to the case of a
non-relativistic rotator, whose Hamiltonian is invariant with respect to the
left transformations (external symmetries), being, however, generally
non-invariant under some (or even all) of the right transformations, depending
on the degree of violation of an internal symmetry. RWE and the corresponding
Lagrangians must be invariant under the left transformations of the
Poincar\'{e} group (Lorentz transformations), but may be non-invariant under
some of the right transformations, which corresponds to the violation of an
internal symmetry.

Thus, one may attempt to obtain equation with \textit{a broken right (that is,
internal) symmetry} acting by analogy with the case of a non-relativistic
rotator. The Hamiltonian of a non-relativistic rotator is constructed from the
right generators $\hat I_{k}$ of the group of rotations $SU(2)$; usually the
Hamiltonian $H=H(\hat I_{k})$ is considered as the sum of squared right
generators with various coefficients, see (\ref{rot.11}).

A Lagrangian for the relativistic rotator can be chosen as a function of right
generators of the Lorentz group $SL(2,C)$, $L=L(\hat S^R_{\mur\nur})$.
However, our purpose here is to take into account not only rotational but also translational movement,
so we must take account not only of the orientation coordinates $z$, which are used to
construct the operators $\hat S^R_{\mur\nur}$, but also of
the space-time coordinates $x^\mu$.

The simplest opportunity is to consider equations for the eigenvalues
of the generators of right translations (\ref{gen.R}),
\begin{equation}
\hat p^R_\mur f(x,z) = \kappa_\mur f(x,z).
\end{equation}
However, the generators $\hat p^R_\mur = \hat p_\mu v^\mu_{\;\;\mur}$
do not commute with the Casimir operators (\ref{gen.Lcas1}) of the Lorentz group
$SL(2,C)_{right}$, and therefore the corresponding functions $f(x,z)$
cannot be characterized by any fixed Lorentz characteristics $j_1,j_2$. Furthermore,
the explicit form of the operators $\hat p^R_\mur$ implies that the functions $f(x,z)$
must contain arbitrarily large powers in $z$ and the corresponding representation of the
Lorentz group must be infinite-dimensional.
The above-said is also valid for the left-invariant operators
\begin{equation}
\hat p_\mu\hat S^{\mu\nu}v_\nu^{\;\;\mur}\;.
\end{equation}

Nevertheless, there exist another possible approach, which is based on the use
of the operators\footnote{A construction of the basic types of RWE as
equations for the eigenvalues of certain sets of commuting operators acting in
the space of scalar functions on the Poincar\'e group was carried out in our
paper \cite{GitSh01}. However, we did not present a systematic consideration
of the properties of these equations with respect to the right
transformations.}

\begin{eqnarray}
\hat \Gamma^{\mu \nur} &=& \hat V_{12}\mathstrut^{\mu \nur}+\hat V_{21}\mathstrut^{\mu \nur} =
\frac{1}{2}\left( \bar \sigma ^{\mu{\dot\alpha}\beta} \sigma^{\nur}_{ \;\; \betar\dot\alphar}
\cc z_{{\dot\alpha}}^{\;\;\dot\alphar}\partial_{\beta}^{\;\;\betar} +
\sigma^{\mu}_{\;\; \beta{\dot\alpha}} \bar \sigma ^{\nur\dot\alphar\betar}
z^{\beta}_{\;\;\betar}\partial^{\dot\alpha}_{\;\;\dot\alphar}\right),
\\
\uGamma^{\mu \nur} &=& \hat V_{12}\mathstrut^{\mu \nur}-\hat V_{21}\mathstrut^{\mu \nur} =
\frac{1}{2}\left( \bar \sigma ^{\mu{\dot\alpha}\beta} \sigma^{\nur}_{ \;\; \betar\dot\alphar}
\cc z_{{\dot\alpha}}^{\;\;\dot\alphar}\partial_{\beta}^{\;\;\betar} -
\sigma^{\mu}_{\;\; \beta{\dot\alpha}} \bar \sigma ^{\nur\dot\alphar\betar}
z^{\beta}_{\;\;\betar}\partial^{\dot\alpha}_{\;\;\dot\alphar}\right),
\end{eqnarray}
(with one external (left) and one internal (right) index) satisfying the commutation relations
\begin{eqnarray}
&&[\hat\Gamma^{\mu\mur},\hat\Gamma^{\nu\nur}]=-i(\hat S^{\mu\nu}\eta^{\mur\nur}+\hat S^{\mur\nur}\eta^{\mu\nu}),
\label{comg2}
\\
&&[\hat S^{\lambda\mu},\hat\Gamma^{\nu\nur}]
=i(\eta^{\mu\nu}\hat \Gamma^{\lambda\nur} - \eta^{\lambda\nu}\hat \Gamma^{\mu\nur} ), \qquad
[\hat S^{{\underline{l}}\mur},\hat\Gamma^{\nu\nur}]
=i(\eta^{\mur\nur}\hat \Gamma^{\nu{\underline{l}}} - \eta^{{\underline{l}}\nur}\hat \Gamma^{\nu\mur} ),
\label{comg1}
\\
&&[\uGamma^{\mu\mur},\uGamma^{\nu\nur}]=i(\hat S^{\mu\nu}\eta^{\mur\nur}+\hat S^{\mur\nur}\eta^{\mu\nu}),
\label{comg_2}
\\
&&[\hat S^{\lambda\mu},\uGamma^{\nu\nur}]
=i(\eta^{\mu\nu}\uGamma^{\lambda\nur} - \eta^{\lambda\nu}\uGamma^{\mu\nur} ), \qquad
[\hat S^{{\underline{l}}\mur},\uGamma^{\nu\nur}]
=i(\eta^{\mur\nur}\uGamma^{\nu{\underline{l}}} - \eta^{{\underline{l}}\nur}\uGamma^{\nu\mur} ),
\label{comg_1}
\\
&&
[\hat\Gamma^5,\hat \Gamma^{\mu\mur}]=-\uGamma^{\mu\mur},  \quad
[\hat\Gamma^5,\uGamma^{\mu\mur}]=\hat \Gamma^{\mu\mur}.
\end{eqnarray}
For space reflection, in accordance with (\ref{xz}), we have
\begin{equation}\label{PGamma}
P:\qquad \hat \Gamma^{\mu \nur} \to (-1)^{\mu + \nur}\hat \Gamma^{\mu \nur},
\qquad \uGamma^{\mu \nur} \to -(-1)^{\mu + \nur}\uGamma^{\mu \nur}.
\end{equation}
The operators $\hat \Gamma^{\mu \nur}$ and $\uGamma^{\mu \nur}$ relate the irreps of the Lorentz group $T_{[j_1,j_2]}$ with the irreps
$T_{[j_1\!+\!1\;j_2\!-\!1]}$ and $T_{[j_1\!-\!1\;j_2\!+\!1]}$.
In the irreps $T_{[j_1j_2]}$ the eigenvalue of the chirality operator (\ref{gen.gamma5}) is $\Gamma^5=j_1-j_2$,
and therefore they join states of different chirality and with a fixed sum $j_1+j_2$,
$|\Gamma^5|\leq j_1+j_2$ into one multiplet.

Using contractions with respect to the left index $\hat p_\mu \hat \Gamma^{\mu \nur}$ and $\hat p_\mu \uGamma^{\mu \nur}$,
we have 8 \textit{left-invariant} equations:
\begin{eqnarray} \label{g-eq}
( \hat p_\mu \hat \Gamma^{\mu \nur} - \varkappa^n )f(x,z) &=& 0,
\\   \label{g_-eq}
( \hat p_\mu \uGamma^{\mu \nur} - \underline\varkappa^n )f(x,z) &=& 0.
\end{eqnarray}
In accordance with (\ref{PGamma}), it is only the operators $\hat p_\mu \hat \Gamma^{\mu \underline{0}}$
and $\hat p_\mu \uGamma^{\mu i}$, $i=1,2,3$
that are invariant under space reflection, and the associated four equations possess solutions
with a definite inner parity.

The commutation relations (\ref{comg1}) and (\ref{comg_1}) entail that the left-invariant equations
under consideration also possess some of the right symmetries.
If the equations related to the operators $\hat\Gamma^{\mu \underline{0}}$ and $\uGamma^{\mu \underline{0}}$
possess a symmetry with respect to $SU(2)={\rm Spin}(3)\in SL(2,C)_{right}$, then the equations related to
the operators $\hat\Gamma^{\mu \underline{i}}$ and $\uGamma^{\mu \underline{i}}$,
$i=1,2,3$, possess a symmetry with respect to three different subgroups $SU(1,1)={\rm Spin}(2,1)\in SL(2,C)_{right}$.

Let us make a more detailed analysis of the equation
\begin{equation}\label{2dir}
( \hat p_\mu \hat \Gamma^{\mu \underline{0}} - \varkappa^{\underline{0}} )f(x,z)=0,
\end{equation}
related to the temporal component of the right 4-vector $\hat p_\mu \hat \Gamma^{\mu \nur}$,
\begin{eqnarray}
\nonumber
2\hat\Gamma^{\mu \underline{0}}
& = &
\bar \sigma ^{\mu{\dot\alpha}\beta} (\cc z_{{\dot\alpha}}^{\;\;\dot{\underline{1}}}\partial_{\beta}^{\;\;\underline{1}}
+ \cc z_{{\dot\alpha}}^{\;\;\dot {\underline{2}}}\partial_{\beta}^{\;\;\underline{2}})
+ \sigma^{\mu}_{\;\; \beta{\dot\alpha}} ( z^{\beta}_{\;\;\underline{1}}\partial^{\dot \alpha}_{\;\;\dot{\underline{1}}}
+ z^{\beta}_{\;\;\underline{2}}\partial^{\dot \alpha}_{\;\;\dot {\underline{2}}}) \\
& = &
(\bar \sigma ^{\mu{\dot\alpha}\beta} \cc{\underline z}_{{\dot\alpha}}\partial/\partial z^{\beta} +
\sigma^{\mu}_{\;\; \beta{\dot\alpha}} z^{\beta}\partial/\partial \cc {\underline z}_{\dot \alpha})
+
(\bar \sigma ^{\mu{\dot\alpha}\beta} \cc z_{{\dot\alpha}}\partial/\partial {\underline z}^{\beta} +
\sigma^{\mu}_{\;\; \beta{\dot\alpha}} {\underline z}^{\beta}\partial/\partial \cc z_{\dot \alpha}).
\end{eqnarray}

It is easy to see that the subspaces $V_{-}$ (the functions
$f(x^\mu,z^\alpha_{\;\;\underline{1}},\cc z^{\dot \alpha}_{\;\;\dot {\underline{2}}})$)
and $V_{+}$ (the functions
$f(x^\mu,z^\alpha_{\;\;\underline{2}},\cc z^{\dot \alpha}_{\;\;\dot {\underline{1}}})$)
are invariant not only with respect to the operators $\hat p_\mu \hat \Gamma^{\mu \underline{0}}$, but also for the
operators of parity $P$ and time reversal $T$.
The action of the operators $\hat\Gamma^{\mu \underline{0}}$ on a row $(z^1\, z^2\, \cc{\underline z}_{\dot 1}\, \cc{\underline z}_{\dot 2})$
amounts to the multiplication of the latter by the 4x4 matrices $\gamma^\mu/2$,
\begin{equation*}
\gamma^\mu = \left( \begin{array}{cc}
 0 & \sigma^\mu \\
 \bar\sigma^\mu & 0
  \\
\end{array} \right).
\end{equation*}
Substituting the $z$-linear functions corresponding to spin 1/2 from the subspace $V_{-}$,
\begin{equation}
f_{D}(x,z)
=\chi_\alpha (x)z^\alpha + \cc\psi^{{\dot\alpha}} (x)\cc{\underline z}_{{\dot\alpha}}=Z_D\Psi_D(x),
\quad Z_D=(z^\alpha \; \cc{\underline z}_{{\dot\alpha}} ), \quad
\Psi_D(x)={\binom{\chi_\alpha (x)}{\cc \psi^{{\dot\alpha}}(x)} },
\label{fDir}
\end{equation}
into equation (\ref{2dir}) and making a comparison of the coefficients at
$z^\alpha$ and $\cc{\underline z}_{\dot\alpha}$ in the left- and right-hand
parts, we obtain the Dirac equation
\begin{equation}  \label{rwe.dir}
(\hat p_\mu \gamma^\mu - \varkappa)\Psi_D(x)=0, \quad
\end{equation}
where $\varkappa=2\varkappa^{\underline{0}}$.
The action of the chirality operator (\ref{gen.gamma5}) amounts to the multiplication by $\gamma^5/2$,
$\gamma^5=\mathrm{diag} \{\sigma^0,-\sigma^0\}$: $\hat\Gamma^5 f_{D}(x,z)= \frac{1}{2} Z_D \gamma^5 \Psi_D(x)$.

For the states with a definite momentum, (\ref{rwe.dir}) is a set of homogenous equations
$(p_\mu \gamma^\mu - \varkappa)\Psi_D(x)=0$, the existence of whose non-trivial solutions
demands that its determinant, equal to $(p^2-\varkappa^2)^2$, must turn to zero, which
implies $\varkappa=\varepsilon m$, where $\varepsilon=\pm 1$.

A charge-conjugate state corresponds to a complex-conjugate function from the subspace $V_{+}$,
\[
\cc f_{D}(x,z) = - \psi_{\alpha}(x){\underline z}^\alpha
                   - \cc\chi^{\dot\alpha}(x)\cc z_{\dot\alpha},
\]
(the minus sign arises due to spinor anticommutation,
$\psi_\alpha z^\alpha= -z_\alpha \psi^\alpha$) or, equivalently, in the matrix form
\begin{equation} \label{rwe.18}
Z_D\Psi_D(x)\stackrel{C}{\to}\cc Z_D\cc\Psi_D(x)=\underline Z_D \Psi^c_D(x),
\quad
\Psi^c_D(x)=- \binom{\psi_{\alpha}(x)} {\cc\chi^{\dot\alpha}(x)}
  =i\sigma^2  \binom{\psi^{\alpha}(x)} {-\cc\chi_{\dot\alpha}(x)},
\end{equation}
where $\underline Z_D=({\underline z}^\alpha, \cc z_{\dot\alpha})$ and $Z_D$ have the same
transformation law with respect to $SL(2,C)_{left}$.
We have thus obtained various scalar functions $f(x,z)$ for the description of
particles and antiparticles, and correspondingly, two Dirac equations at the
same time, for both signs of the charge. This is in good agreement with the
results of the article \cite{Git}, which shows that a consistent quantization
of a classical model of a spinning particle entails as a result exactly the
same (charge-symmetrical) quantum mechanics. It is completely equivalent to
the one-particle sector of the corresponding quantum field theory.

Consequently, in the case of $z$-linear functions of general form,
\begin{equation} \label{rwe.19}
f(x,z) =  \chi_{\alpha}^{\;\;\alphar}(x)z^\alpha_{\;\;\alphar}
              +\cc\psi^{\dot\alpha}_{\;\;\dot\betar}(x)\cc z_{\dot\alpha}^{\;\;\dot\betar}
              =(z^\alpha_{\;\;\underline{1}}\,  \cc z_{\dot\alpha}^{\;\;\underline{1}}\,
                z^\alpha_{\;\;\underline{2}}\, \cc z_{\dot\alpha}^{\;\;\underline{2}})
                (\chi_{\alpha}^{\;\;\underline{1}}\, \cc\psi^{\dot\alpha}_{\;\;\dot{\underline{1}}}\,
                \chi_{\alpha}^{\;\;\underline{2}}\, \cc\psi^{\dot\alpha}_{\;\;\dot{\underline{2}}} )^T,
\end{equation}
equation (\ref{2dir}) can be represented in the matrix form
\begin{equation} \label{rwe.20}
p_\mu\gamma^{\mu \underline{0}}\Psi(x)=\varkappa \Psi(x), \qquad \gamma^{\mu \underline{0}}=\mathrm{diag} (\gamma^{\mu}, \gamma^{\mu}),
\end{equation}
where $\Psi(x)=(\chi_{\alpha}^{\;\;\underline{1}}\, \cc\psi^{\dot\alpha}_{\;\;\dot{\underline{1}}}\,
                \chi_{\alpha}^{\;\;\underline{2}}\, \cc\psi^{\dot\alpha}_{\;\;\dot{\underline{2}}} )^T$
is a column of $8$ coefficients at
$z^{\alpha}_{\;\;\underline{1}}\,, \cc z_{\dot\alpha}^{\;\;\dot{\underline{1}}}\,,
z^{\alpha}_{\;\;\underline{2}}\,,\cc z_{\dot\alpha}^{\;\;\dot{\underline{2}}}$,
and, therefore, in the case of $z$-linear functions it is equivalent to a pair
of Dirac equations with the same mass, that transform into each another at charge conjugation $C$.
This is an eight-component equation and the corresponding Lagrangian
\begin{equation} \label{rwe.20a}
\mathcal{L}= i\overline{\Psi}\gamma^{\mu\underline{0}}\partial^\mu\Psi - m\overline{\Psi}\Psi, \quad
\overline{\Psi} = \Psi^\dagger \gamma^{0\underline{0}},
\end{equation}
is invariant under right rotations, as distinct from right boosts;
the operators of right rotations, owing to (\ref{comg1}), commute with $\hat \Gamma^{\mu \underline{0}}$.
However, each of these two Dirac equations is separately invariant only with respect to
right rotations generated by $\hat S^R_3$, since in the case of the other two generators ($\hat S^R_1$ and $\hat S^R_2$)
the spaces of functions $V_{-}$ and $V_{+}$ are not invariant.
This means that in the series --
equation (\ref{casimir-p2}) for the Casimir operator $\hat p^2$ -- equation (\ref{2dir}) -- Dirac equation (\ref{rwe.dir})
-- there occurs a reduction of the right (inner) symmetries:
\begin{equation}\label{rwe.symm}
SL(2,C)_{right}\times U(1) \to SU(2)\times U(1) \to U(1)\times U(1).
\end{equation}

Let us emphasize the fact the introduction of the ``Dirac'' mass with the help of
the left-invariant equation (\ref{2dir}) is related with a violation of
the symmetry with respect to right boosts.
The functions that satisfy the equation are characterized by a definite parity
and present a superposition of states with different chirality.
A priori, judging from a purely group-theoretical viewpoint, there is no obstacle
for an introduction of mass without having to impose equation (\ref{2dir}),
which would entail the existence of massive chiral fermions, satisfying only
the equations (\ref{casimir-p2}), (\ref{casimir-w2})
for the Casimir operators. In this case, the right symmetry remains unbroken.

Let us now consider equations (\ref{g-eq}) at $n=1,2,3$, related to
the spatial components of the right 4-vector
$\hat p_\mu \hat \Gamma^{\mu \nur}$.

The action of the operators $\hat\Gamma^{\mu 3}$,
\begin{eqnarray} \nonumber
2\hat\Gamma^{\mu 3}
& = &
\bar \sigma ^{\mu{\dot\alpha}\beta} (\cc z_{{\dot\alpha}}^{\;\;\dot{\underline{1}}}\partial_{\beta}^{\;\;\underline{1}}
- \cc z_{{\dot\alpha}}^{\;\;\dot {\underline{2}}}\partial_{\beta}^{\;\;\underline{2}})
- \sigma^{\mu}_{\;\; \beta{\dot\alpha}} ( z^{\beta}_{\;\;\underline{1}}\partial^{\dot \alpha}_{\;\;\dot{\underline{1}}}
- z^{\beta}_{\;\;\underline{2}}\partial^{\dot \alpha}_{\;\;\dot {\underline{2}}}) \\
& = &
(\bar \sigma ^{\mu{\dot\alpha}\beta} \cc{\underline z}_{{\dot\alpha}}\partial/\partial z^{\beta} -
\sigma^{\mu}_{\;\; \beta{\dot\alpha}} z^{\beta}\partial/\partial \cc{\underline z}_{\dot \alpha})
-
(\bar \sigma ^{\mu{\dot\alpha}\beta} \cc z_{{\dot\alpha}}\partial/\partial {\underline z}^{\beta} -
\sigma^{\mu}_{\;\; \beta{\dot\alpha}} {\underline z}^{\beta}\partial/\partial \cc z_{\dot \alpha}),
\label{rwe.21}
\end{eqnarray}
on a row $(z^1\, z^2\, \cc{\underline z}_{\dot 1}\, \cc{\underline z}_{\dot 2})$
amounts to its multiplication by the 4x4 matrices
\begin{equation} \label{rwe.22}
\tilde\gamma^{\mu}=\left( \begin{array}{cc}
 0 & -\sigma^\mu \\
 \bar\sigma^\mu & 0
  \\
\end{array} \right).
\end{equation}
Substituting $f(x,z) =\chi_\alpha (x)z^\alpha + \cc\psi^{{\dot\alpha}} (x)\cc{\underline z}_{{\dot\alpha}}=(z^\alpha\; \cc{\underline z}_{{\dot\alpha}})\Psi(x)$
into equation (\ref{g-eq}) at $n=3$, we obtain $(\hat p_\mu \tilde \gamma^{\mu} - \varkappa)\Psi(x)=0$,
where $\varkappa=2\varkappa^3$.

Multiplying $(\hat p_\mu \tilde \gamma^{\mu} - \varkappa)$ by $(\hat p_\mu \tilde\gamma^\mu + \varkappa)$,
we find that $\hat p_\mu^2\Psi_(x) = - \varkappa^2\Psi_(x)$, whence one can deduce that
the corresponding equations are related to tachyons.
A similar conclusion can be reached by considering states with a definite momentum. In this case,
we have a set of homogenous equations $(p_\mu \tilde\gamma^\mu - \varkappa)\Psi(x)=0$,
the existence of whose non-trivial solutions demands that its determinant, equal to $(p^2+\varkappa^2)^2$,
must turn to zero.

For the operators $\hat\Gamma^{\mu \underline{1}}$ and $\hat\Gamma^{\mu \underline{2}}$,
\begin{eqnarray}
\nonumber
2\hat\Gamma^{\mu \underline{1}}
& = &
\bar \sigma ^{\mu{\dot\alpha}\beta} (\cc z_{{\dot\alpha}}^{\;\;\dot {\underline{2}}}\partial_{\beta}^{\;\;\underline{1}}
+ \cc z_{{\dot\alpha}}^{\;\;\dot{\underline{1}}} \partial_{\beta}^{\;\;\underline{2}})
- \sigma^{\mu}_{\;\; \beta{\dot\alpha}} ( z^{\beta}_{\;\;\underline{1}}\partial^{\dot \alpha}_{\;\;\dot {\underline{2}}}
+ z^{\beta}_{\;\;\underline{2}}\partial^{\dot \alpha}_{\;\;\dot {\underline{1}}}) \\
& = &
(\bar \sigma ^{\mu{\dot\alpha}\beta} \cc z_{{\dot\alpha}}\partial/\partial z^{\beta} -
\sigma^{\mu}_{\;\; \beta{\dot\alpha}} z^{\beta}\partial/\partial \cc z_{\dot \alpha})
+
(\bar \sigma ^{\mu{\dot\alpha}\beta} \cc {\underline z}_{{\dot\alpha}}\partial/\partial {\underline z}^{\beta} -
\sigma^{\mu}_{\;\; \beta{\dot\alpha}} {\underline z}^{\beta}\partial/\partial \cc {\underline z}_{\dot \alpha}),
\\
\nonumber
2\hat\Gamma^{\mu \underline{2}}
& = &
i\bar \sigma ^{\mu{\dot\alpha}\beta} (\cc z_{{\dot\alpha}}^{\;\;\dot {\underline{2}}}\partial_{\beta}^{\;\;\underline{1}}
- \cc z_{{\dot\alpha}}^{\;\;\dot {\underline{1}}}\partial_{\beta}^{\;\;\underline{2}})
- i\sigma^{\mu}_{\;\; \beta{\dot\alpha}} ( z^{\beta}_{\;\;\underline{1}}\partial^{\dot \alpha}_{\;\;\dot {\underline{2}}}
- z^{\beta}_{\;\;\underline{2}}\partial^{\dot \alpha}_{\;\;\dot {\underline{1}}}) \\
& = &
i(\bar \sigma ^{\mu{\dot\alpha}\beta} \cc z_{{\dot\alpha}}\partial/\partial z^{\beta} -
\sigma^{\mu}_{\;\; \beta{\dot\alpha}} z^{\beta}\partial/\partial \cc z_{\dot \alpha})
-
i(\bar \sigma ^{\mu{\dot\alpha}\beta} \cc {\underline z}_{{\dot\alpha}}\partial/\partial {\underline z}^{\beta} -
\sigma^{\mu}_{\;\; \beta{\dot\alpha}} {\underline z}^{\beta}\partial/\partial \cc {\underline z}_{\dot \alpha}).
\end{eqnarray}
the invariant subspaces are given by the functions $f(x,{\underline z},\cc{\underline z})$ and $f(x, z,\cc z)$.

For $\bf z$-linear functions $f(x, z)$, the left-invariant equations,
related to the spatial components of the right 4-vector $\hat p_\mu \hat \Gamma^{\mu \nur}$,
can be presented in the matrix form as follows:
\begin{eqnarray} \label{rwe.24}
&&\hat p_\mu\gamma^{\mu \nur}\Psi(x)=2\varkappa^\nur \Psi(x),
\\ \label{rwe.25}
&&\gamma^{\mu \underline{1}}=\left( \begin{array}{cc}
 0 & \tilde\gamma^\mu \\
 \tilde\gamma^\mu & 0
\end{array} \right), \;
\gamma^{\mu \underline{2}}=\left( \begin{array}{cc}
 0 & -i \tilde\gamma^\mu \\
 i \tilde\gamma^\mu & 0
\end{array} \right), \;
\gamma^{\mu 3}=\left( \begin{array}{cc}
 \tilde\gamma^\mu & 0\\
0 & - \tilde\gamma^\mu
\end{array} \right),
\nonumber
\end{eqnarray}
where $\Psi^n(x)$ is an 8-component column being composed of the coefficients at $z,\cc {\underline z}, {\underline z},\cc z$
in the power expansion of the function $f(x,z)$. These equations describe tachyons.
The matrices $\gamma^{\mu\nur}$ obey the conditions
\begin{equation} \label{rwe.26}
[\gamma^{\mu\nur},\gamma^{\nu\nur}]_+ = 2\eta^{\nur\nur}\eta^{\mu\nu}, \quad
(\gamma^{\mu\nur})^{\dagger}= (-1)^{\delta_{\mu 0}+\delta_{\nur \underline{0}}}\gamma^{\mu\nur}.
\end{equation}

Let us now present the expressions for the operators $\uGamma^{\mu\mur}$
that enter equations (\ref{g_-eq}), namely:
\begin{eqnarray}
2\uGamma^{\mu \underline{0}}
& = &
\bar \sigma ^{\mu{\dot\alpha}\beta} (\cc z_{{\dot\alpha}}^{\;\;\dot {\underline{1}}}\partial_{\beta}^{\;\;\underline{1}}
- \cc z_{{\dot\alpha}}^{\;\;\dot {\underline{2}}}\partial_{\beta}^{\;\;\underline{2}})
- \sigma^{\mu}_{\;\; \beta{\dot\alpha}} ( z^{\beta}_{\;\;\underline{1}}\partial^{\dot \alpha}_{\;\;\dot {\underline{1}}}
- z^{\beta}_{\;\;\underline{2}}\partial^{\dot \alpha}_{\;\;\dot {\underline{2}}})
\nonumber\\
& = &
(\bar \sigma ^{\mu{\dot\alpha}\beta} \cc{\underline z}_{{\dot\alpha}}\partial/\partial z^{\beta} -
\sigma^{\mu}_{\;\; \beta{\dot\alpha}} z^{\beta}\partial/\partial \cc {\underline z}_{\dot \alpha})
+
(\bar \sigma ^{\mu{\dot\alpha}\beta} \cc z_{{\dot\alpha}}\partial/\partial {\underline z}^{\beta} -
\sigma^{\mu}_{\;\; \beta{\dot\alpha}} {\underline z}^{\beta}\partial/\partial \cc z_{\dot \alpha}),
\nonumber
\\
2\uGamma^{\mu 3}
& = &
\bar \sigma ^{\mu{\dot\alpha}\beta} (\cc z_{{\dot\alpha}}^{\;\;\dot {\underline{1}}}\partial_{\beta}^{\;\;\underline{1}}
- \cc z_{{\dot\alpha}}^{\;\;\dot {\underline{2}}}\partial_{\beta}^{\;\;\underline{2}})
+ \sigma^{\mu}_{\;\; \beta{\dot\alpha}} ( z^{\beta}_{\;\;\underline{1}}\partial^{\dot \alpha}_{\;\;\dot {\underline{1}}}
- z^{\beta}_{\;\;\underline{2}}\partial^{\dot \alpha}_{\;\;\dot {\underline{2}}})
\nonumber\\
& = &
(\bar \sigma ^{\mu{\dot\alpha}\beta} \cc{\underline z}_{{\dot\alpha}}\partial/\partial z^{\beta} +
\sigma^{\mu}_{\;\; \beta{\dot\alpha}} z^{\beta}\partial/\partial \cc{\underline z}_{\dot \alpha})
-
(\bar \sigma ^{\mu{\dot\alpha}\beta} \cc z_{{\dot\alpha}}\partial/\partial {\underline z}^{\beta} +
\sigma^{\mu}_{\;\; \beta{\dot\alpha}} {\underline z}^{\beta}\partial/\partial \cc z_{\dot \alpha}),
\nonumber
\\
2\uGamma^{\mu \underline{1}}
& = &
\bar \sigma ^{\mu{\dot\alpha}\beta} (\cc z_{{\dot\alpha}}^{\;\;\dot {\underline{2}}}\partial_{\beta}^{\;\;\underline{1}}
+ \cc z_{{\dot\alpha}}^{\;\;\dot {\underline{1}}}\partial_{\beta}^{\;\;\underline{2}})
+ \sigma^{\mu}_{\;\; \beta{\dot\alpha}} ( z^{\beta}_{\;\;\underline{1}}\partial^{\dot \alpha}_{\;\;\dot {\underline{2}}}
+ z^{\beta}_{\;\;\underline{2}}\partial^{\dot \alpha}_{\;\;\dot {\underline{1}}})
\nonumber\\
& = &
(\bar \sigma ^{\mu{\dot\alpha}\beta} \cc z_{{\dot\alpha}}\partial/\partial z^{\beta} +
\sigma^{\mu}_{\;\; \beta{\dot\alpha}} z^{\beta}\partial/\partial \cc z_{\dot \alpha})
+
(\bar \sigma ^{\mu{\dot\alpha}\beta} \cc {\underline z}_{{\dot\alpha}}\partial/\partial {\underline z}^{\beta} +
\sigma^{\mu}_{\;\; \beta{\dot\alpha}} {\underline z}^{\beta}\partial/\partial \cc {\underline z}_{\dot \alpha}),
\nonumber
\\
2\uGamma^{\mu \underline{2}}
& = &
i\bar \sigma ^{\mu{\dot\alpha}\beta} (\cc z_{{\dot\alpha}}^{\;\;\dot {\underline{2}}}\partial_{\beta}^{\;\;\underline{1}}
- \cc z_{{\dot\alpha}}^{\;\;\dot {\underline{1}}}\partial_{\beta}^{\;\;\underline{2}})
+ i\sigma^{\mu}_{\;\; \beta{\dot\alpha}} ( z^{\beta}_{\;\;\underline{1}}\partial^{\dot \alpha}_{\;\;\dot {\underline{2}}}
- z^{\beta}_{\;\;\underline{2}}\partial^{\dot \alpha}_{\;\;\dot {\underline{1}}})
\nonumber\\
& = &
i(\bar \sigma ^{\mu{\dot\alpha}\beta} \cc z_{{\dot\alpha}}\partial/\partial z^{\beta} +
\sigma^{\mu}_{\;\; \beta{\dot\alpha}} z^{\beta}\partial/\partial \cc z_{\dot \alpha})
-
i(\bar \sigma ^{\mu{\dot\alpha}\beta} \cc {\underline z}_{{\dot\alpha}}\partial/\partial {\underline z}^{\beta} +
\sigma^{\mu}_{\;\; \beta{\dot\alpha}} {\underline z}^{\beta}\partial/\partial \cc {\underline z}_{\dot \alpha}).
\end{eqnarray}

In the matrix form, equations (\ref{g_-eq}) for the $z$-linear functions
are written as follows:
\begin{align}  \label{rwe.28}
&\hat p_\mu\underline\gamma^{\mu \nur}\Psi^n(x)=2\underline\varkappa^\nur \Psi^n(x),
\\  \label{rwe.29}
&\underline\gamma^{\mu \underline{0}}=\!\left( \begin{array}{cc}
 \tilde\gamma^\mu\! & 0\\
 0 & \tilde\gamma^\mu\!
\end{array} \right)\!, \;
\underline\gamma^{\mu \underline{1}}=\!\left( \begin{array}{cc}
 0 & \gamma^\mu\! \\
 \gamma^\mu\! & 0
\end{array} \right)\!, \;
\underline\gamma^{\mu \underline{2}}=\!\left( \begin{array}{cc}
 0 & -i \gamma^\mu\!\! \\
 i \gamma^\mu\!\! & 0
\end{array} \right)\!, \;
\underline\gamma^{\mu 3}=\!\left( \begin{array}{cc}
 \gamma^\mu\!\!\! & 0\\
0 & - \gamma^\mu\!\!\!
\end{array} \right)\!.
\nonumber
\end{align}
The matrices $\underline\gamma^{\mu\nur}$ obey the conditions
\begin{equation} \label{rwe.30}
[\underline\gamma^{\mu\mur},\underline\gamma^{\nu\mur}]_+ = -\eta^{\mur\mur}\eta^{\mu\nu}, \quad
(\underline\gamma^{\mu\mur})^{\dagger}= -(-1)^{\delta_{\mu 0}+\delta_{\mur \underline{0}}}\underline\gamma^{\mu\mur}.
\end{equation}
It is easy to see that the equation related to $\underline\gamma^{\mu \underline{0}}$ describe tachyons
and the remaining three equations -- usual particles (bradyons) of spin 1/2.
For the latter equations, the Hermitian-conjugate ones (\ref{rwe.28}), owing to (\ref{rwe.30}), we present
in the form
\begin{equation} \label{rwe.31}
\overline{\Psi}(i\underline\gamma^{\mu\underline{k}}\overleftarrow{\partial}_{\mu}+2\varkappa )=0, \quad
\overline{\Psi}={\Psi}^\dagger\underline\gamma^{0\underline{k}},
\end{equation}
where $2\varkappa=\pm m$, which allows us to present the corresponding currents
$j^\mu=\overline{\Psi}\underline\gamma^{\mu\underline{k}}\Psi$ and Lagrangians,
\begin{equation} \label{rwe.32}
\mathcal{L}^{\underline{k}}= i\overline{\Psi}\underline\gamma^{\mu\underline{k}}\partial_\mu\Psi - 2\varkappa\overline{\Psi}\Psi, \quad
\overline{\Psi} = \Psi^\dagger \underline\gamma^{0\underline{k}}.
\end{equation}
Notice that the eight-component equation related to $\underline\gamma^{\mu \underline{3}}$
splits into two Dirac equations with opposite signs of the mass term. As the equation (\ref{rwe.20}),
it is invariant under rotations generated by $\hat S^3_R$; it is, however, not invariant
under rotations generated by $\hat S^1_R$ and $\hat S^2_R$.

A special role played by the Dirac equation in relativistic theory can be largely explained
by the fact that it actually contains all the four types of spinors: two of them in a manifest form
and the other two are obtained as a result of the complex conjugation of the equation, as
well as by the consideration of both signs of the mass term.
The study of the properties of the Dirac equation allows one to establish, in particular,
the properties of discrete transformations, as well as to introduce chirality, being
the quantum number corresponding to the operator $\gamma^5$.

One can also do the contrary, by examining both the left and right
generators of the Poincar\'e group, as well as the corresponding
quantum numbers. This allows one not only to give a consistent
description of discrete transformations, chirality, particles and
antiparticles without the aid of RWE, but also to give an exact
group-theory formulation of the conditions for which the equation is
valid (in other words, to give a group-theory deduction of the Dirac
equation, as was done in \cite{BucGiS02}). Such a formulation
enables one, in particular, to establish the relation of the
mass-term sign with the characteristics of the extended Poincar\'e
group.

Let us return to the equation (\ref{2dir}) for the operator $\hat\Gamma^{\mu \underline{0}}$,
for $z$-linear functions from the subspaces; this equation is equivalent to a pair
of Dirac equations. Polynomials of degree $2s$ from the subspaces $V_{-}$ and $V_{+}$ are
eigenfunctions of the operator $\hat S^3_R$ with the eigenvalues $s$ and $-s$.

In the case of $z$-quadratic functions from the subspace $V_{-}$,
\begin{align} \label{rwe.DK}
&f(x,z)=
\chi_{\alpha\beta}(x) z^\alpha z^\beta
   +\phi_{\alpha}^{\;\;{\dot\beta}}(x) z^{\alpha} \cc{\underline z}_{\dot\beta}
   +\psi^{\dot\alpha{\dot\beta}}(x) \cc{\underline z}_{\dot\alpha} \cc{\underline z}_{\dot\beta}
   =\Phi_\mu (x) q^\mu + \frac{1}{2} F_{\mu\nu}(x) q^{\mu\nu} ,
\\
&q^\mu=\frac{1}{2}{\sigma^{\mu}}_{\alpha{\dot\beta}}z^\alpha \cc{\underline z}^{{\dot\beta}}, \quad
q_\mu q^\mu=0, \quad
q_{\mu\nu}=-q_{\nu\mu}=\frac{1}{2} \left((\sigma_{\mu\nu})_{\alpha\beta}z^\alpha z^\beta+
(\bar\sigma_{\mu\nu})_{{\dot\alpha}{\dot\beta}} \cc{\underline z}^{{\dot\alpha}}\cc{\underline z}^{{\dot\beta}}\right),
\nonumber\\
&\Phi_\mu(x)= -\bar\sigma_\mu^{\;\;{\dot\beta}\alpha} \phi_{\alpha{\dot\beta}}(x), \quad
F_{\mu\nu}(x)=-2\left( (\sigma_{\mu\nu})_{\alpha\beta}\chi^{\alpha\beta}(x)+
(\bar\sigma_{\mu\nu})_{{\dot\alpha}{\dot\beta}}\psi^{{\dot\alpha}{\dot\beta}}(x)\right),
\nonumber
\end{align}
or in the case of their conjugates from $V_{+}$,
a similar analysis leads to the 10-component equations of Duffin-Kemmer for spin 1 \cite{GitSh01}.

Substituting polynomials of degree $2s$ from the subspaces $V_{-}$ or $V_{+}$ into (\ref{2dir}),
one can obtain a matrix equation of the form
\begin{equation} \label{rwe.33}
(\alpha^\mu \hat p_\mu - \varkappa )\psi=0, \quad
[S^{\lambda\mu},\alpha^\nu]
=i(\eta^{\mu\nu}\alpha^\lambda - \eta^{\lambda\nu}\alpha^\mu ), \quad
[\alpha^\mu ,\alpha^\nu]= S^{\mu\nu}.
\end{equation}
The commutation relations for the matrices follow from the commutation relations
for differential operators, according to which $\hat S^{\mu\nu}$ and $\hat \Gamma^{\mu\underline{0}}$
obey the commutation relations of $SO(3,2)$.
Finite-component equations of the form (\ref{rwe.33})
are known as the Bhabha equations, although for the first time
they were systematically considered by Lubanski.
These equations are classified according to the finite-dimensional irreps of $SO(3,2)$.
The case of $f(x,z)\in V_{-}$ (or $V_{+}$) corresponds to the symmetric irreps
$T_{[2s\,0]}$ of $SO(3,2)$.

It can be demonstrated that in case the eigenvalue $\hat p_\mu \hat\Gamma^{\mu \underline{0}}$ equals to $\pm ms$,
where $2s$ is a (fixed by $\hat S^3_R$) polynomial degree, the eigenvalue of the
Lubanski-Pauli operator $\hat W^2$ corresponds to spin $s$ \cite{GitSh01}.
Consequently, in the case of functions from subspaces $V_{-}$ and $V_{+}$ we arrive at the equation
\begin{equation}
(\hat p_\mu \hat\Gamma^{\mu \underline{0}} - \varepsilon ms) f(x,z)=0, \label{rwe.34}
\end{equation}
where $s$ is spin, $\varepsilon=\pm 1$.
As has been already observed, this equation is invariant with respect to
space reflection. A consideration of the rest frame shows \cite{BucGiS02}
that in equation (\ref{rwe.34}), as well as in the Dirac equation in (\ref{rwe.dir}),
\begin{equation*}
\varepsilon=\eta\sign p_0 \sign S_3^R,
\end{equation*}
i.e. the mass-term sign is the product of the signs of parity, energy and projection $S_3^R$.

\section*{Concluding remarks}

We have examined a field $f(x,z)$ on the Poincar\'{e} group, depending on ten
real-valued parameters. Four of the parameters $x^{\mu }$ correspond to
position, while six parameters correspond to orientation, which is
convenient to describe by the elements of a complex matrix $Z\in
SL(2,C)={\rm Spin}(3,1)$. As distinct from fields in homogenous spaces, in
particular, the Minkowski space $M(3,1)/{\rm Spin}(3,1)$, a field on the entire
group admits two types of transformations -- left and right ones,
corresponding to a change of the laboratory (space-fixed) and local
(body-fixed) reference frames. These transformations commute with each
other. Correspondingly, the variables
$z=\{z_{\;\;\betar}^{\alpha},\,\cc z_{\;\;\dot\betar}^{\dot{\alpha}}\}$,
describing the orientation, have two kinds of indices -- the usual ones
(left, laboratory, indices) and underlined ones (right, local, indices).

For an analysis of physical aspects, we use a close mathematical analogy
with the theory of a three-dimensional non-relativistic rotator, constructed
in the space of functions depending only on three-dimensional orientation
(i.e., functions on the group $SO(3)$), which provides one with a common and
intuitively clear interpretation.

The scalar field $f(x,z)$ contains the fields of all spins and presents a
generating function of usual spin-tensor fields, the latter being the
coefficients of expansion in the powers of the complex-valued variables
$z_{\;\;\betar}^{\alpha }$, $\cc z_{\;\;\dot\betar}^{\dot{\alpha}}$.
Field classification is based on the use of the maximal set of
commuting operators (two Casimir operators and four left and right generators
each, the total being ten, equal to the number of group parameters).

We repeat ones more that in the conventional description of
relativistic particles (in $3+1$ dim.) in terms of spin-tensor
fields, based, in particular, on a classification of Poincar\'{e}
and Lorentz group representations, there appear 8 particle
characteristics (quantum numbers): 2 numbers $j_{1}$ and $j_{2}$
that label Lorentz group representations and 6 numbers related to
the Poincar\'{e} group, those are the mass $m$, spin $s$ (Casimirs
of the group), and 4 numbers, which are eigenvalues of some
combinations of left generators of the group. In particular, the
latter 4 numbers can be some components of the momentum and a spin
projection. The proposed description of relativistic orientable
objects is based on a classification of group representations of
transformation of both s.r.f. and b.r.f., and the orientable
object is characterized already by 10 quantum numbers. These
numbers are related to a maximal set of commuting operators
(\ref{gen.31set}). Here, in addition to the 2 Casimirs and 4 left
generators of the group, we have 4 numbers which are eigenvalues
of some combinations of right generators of the Poincar\'{e}
group. Among the latter, we have 2 Casimirs of the (right) Lorentz
subgroup that fix $j_{1}$ and $j_{2}$. Thus, as compared with the
conventional description, in the proposed one, an oriented object
has two more characteristics. Their physical interpretation is a
subject of an individual investigation.

Field symmetries, defined as transformations of a field into itself, are
divided, from the mathematical viewpoint, into three types: the left
transformations of the Poincar\'e group (Lorentz transformations), right
transformations, and outer automorphisms.

The first type of transformations is regarded as external symmetries (the
symmetries of the embedding space), the second type is regarded as internal
symmetries. Outer automorphisms (reflections $I_{s},I_{t}$, being involutive
automorphisms, and scale transformations) affect both (space-fixed and
body-fixed) reference frames. Charge conjugation corresponds to the complex
conjugation of functions $f(x,z)$.

A scalar field on the Poincar\'{e} group provides an adequate language for a
construction and analysis of the symmetries of RWE. It can be regarded as a
field depending on a 4-vector $x^{\mu }$ and on four types of spinors, two
pairs being related by charge conjugation. These pairs have equal
transformation properties with respect to $SL(2,C)_{left}$ and different
transformation properties with respect to $SL(2,C)_{right}$ and outer
automorphisms. As a consequence of the presence of four types of spinors for
spin $1/2$, there arise eight-component equations corresponding to both signs
of charge, i.e., we thus reproduce the one-particle sector of quantum field
theory.

The equations related to the Casimir operators (Klein-Gordon and
Pauli-Lubanski equations) are invariant with respect not only to the left
but also to the right transformations; however, this is not the case for
other equations. As in the rotator theory, the right (internal) symmetries
may be broken, and the analysis shows that the basic RWE (Dirac,
Duffin-Kemmer equations) possess only some of the right symmetries,
so that only some of the potential right symmetries are realized in Nature.

\section*{Acknowledgement}
The authors are grateful to Profs. S. N. Solodukhin and I. V. Tyutin for useful discussions.
D.M.G. acknowledges the permanent support of FAPESP and CNPq.

\end{document}